\documentclass[useAMS,usenatbib]{mn2e}
\pdfoutput=1 
\usepackage{hyperref}
\usepackage{amsmath}
\usepackage{graphicx}
\usepackage{txfonts}
\usepackage{natbib}

\newcommand{\teff}{T_{\rm{eff}}}

\newcommand{\logg}{\log\,g}
\newcommand{\feh}{\rm{[Fe/H]}}
\newcommand{\aFe}{[\alpha/\rm{Fe}]}
\newcommand{\ergs}{\rm{erg\,s^{-1}\,cm^{-2}\,\AA^{-1}}}
\newcommand{\flux}{\rm{erg\,s^{-1}\,cm^{-2}}}

\newcommand{\kms}{\rm{km\,s^{-1}}}
\newcommand{\fbol}{f_{\rm Bol}}

\newcommand\lta{\mathrel{\hbox{\raise 0.6 ex \hbox{$<$}\kern
                   -1.8 ex\lower .5 ex\hbox{$\sim$}}}}
\newcommand\gta{\mathrel{\hbox{\raise 0.6 ex \hbox{$>$}\kern
                   -1.7 ex\lower .5 ex\hbox{$\sim$}}}}
\newcommand{\RN}[1]{{\small\textup{\uppercase\expandafter{\romannumeral#1}}}}

\title[Synthetic photometry for large scale surveys]{Synthetic Stellar
  Photometry - II. Testing the bolometric flux scale, and tables of bolometric 
  corrections for the Hipparcos/Tycho, Pan-STARRS1, SkyMapper and \emph{JWST}
  systems.}

\author[Casagrande \& VandenBerg]{\parbox{18cm}{
  L.~Casagrande$^{1,2}$\thanks{Email:luca.casagrande@anu.edu.au}, 
  Don A.~VandenBerg$^{3}$}\\%\vspace{0.3cm}\\
  $^1$ Research School of Astronomy and Astrophysics, Mount Stromlo 
  Observatory, The Australian National University, ACT 2611, Australia\\
  $^2$ ARC Centre of Excellence for All Sky Astrophysics in 3 Dimensions
  (ASTRO 3D)\\
  $^3$ Department of Physics \& Astronomy, University of Victoria, P.O.~Box 
  1700 STN CSC, Victoria, BC, V8W 2Y2, Canada}

\begin{document}

\date{Received; accepted}

\maketitle

\begin{abstract}
We use MARCS model atmosphere fluxes to compute synthetic colours, bolometric
corrections and reddening coefficients for the Hipparcos/Tycho, Pan-STARRS1,
SkyMapper and {\it JWST} systems. Tables and interpolation subroutines are
provided to transform isochrones from the theoretical to various observational
planes, to derive bolometric corrections, synthetic colours and
colour-temperature relations at nearly any given point of the HR diagram for
$2600\,\rm{K} \le \teff \le 8000\,\rm{K}$, and different values of reddening
in 85 photometric filters. We use absolute spectrophotometry from the CALSPEC
library to show that bolometric fluxes can be recovered to $\sim 2$ percent
from bolometric corrections in a single band, when input stellar parameters
are well known for FG dwarfs at various metallicities.  This sole source of
uncertainty impacts interferometric $\teff$ to $\simeq 0.5$ percent (or $30$~K
at the solar temperature). Uncertainties are halved when combining bolometric
corrections in more bands, and limited by the fundamental uncertainty of the
current absolute flux scale at $1$ percent. Stars in the RAVE DR5 catalogue are
used to validate the quality of our MARCS synthetic photometry in selected
filters across the optical and infrared range. This investigation shows that
extant MARCS synthetic fluxes are able to reproduce the main features observed
in stellar populations across the Galactic disc.
\end{abstract}


\section{Introduction}

Synthetic stellar photometry is crucial for the translation of theoretical
stellar
quantities, namely effective temperatures ($\teff$), surface gravities
($\logg$), metallicities ($\feh$ often together with $\aFe$) and luminosities
into observables such as
magnitudes and colours in different photometric systems. With the advent of
large scale photometric surveys, this is becoming increasingly important, in
order to allow comparisons of observed stellar populations with theoretical
stellar models, to derive e.g., stellar ages, star-formation histories,
distances etc $\ldots$.
The Gaia mission is providing exquisite distances of stars, which then
translate into precise absolute magnitudes \citep{Gaia16}. However, when
comparing them with stellar models, one of the main limiting factors is the
availability of bolometric corrections (BCs) to translate those magnitudes into
luminosities. 

In this paper we continue our effort to provide reliable synthetic
colours and BCs from the MARCS library of theoretical fluxes
\citep{g08} for different $\teff$, $\logg$, $\feh$ and $\aFe$ combinations .
We follow the approach of \citet[][hereby Paper I]{cv14}, where
detailed information on the MARCS flux library, and the main
concepts of synthetic photometry can be found. Paper I also addresses one of
the main limitation of synthetic colours computed from static 1D model
atmospheres, namely microturbulent velocity, whose effect can be substantial
in the blue and ultraviolet spectral region \citep[see][for an investigation of
synthetic colours from 3D model atmospheres]{chiava,bonifacio}. 
Here, we extend Paper I to
include photometric systems which underpin large scale surveys, and
which will be increasingly used over the next several years. These are the Tycho
system \citep[i.e. the Tycho-2 catalogue, as described in][]{hog2000}
which comprises 2.5 million of the brightest stars in the sky \citep[all part of
the Tycho-Gaia Astrometric Solution,][]{TGAS2015,TGAS2016}; SkyMapper
\citep[a photometric survey of the entire southern sky,][]{wolf}; Pan-STARRS1
\citep[Panoramic Survey Telescope and Rapid Response System in the northern
 hemisphere,][]{pan} and the filters installed on the James Webb Space
Telescope \citep{jwst}. Together with the photometric systems explored in
Paper I (Johnson-Cousins, SDSS, 2MASS, {\it HST}-ACS and {\it HST}-WFC3) our
transformations provide one of the most extended set of homogeneously derived
and tested
BCs and synthetic colours available in the literature.
Expanding from Paper, I we provide an explicit formulation to translate
bolometric corrections in physical fluxes, and use
a number of well calibrated flux standards from the CALSPEC
library\footnote{\href{http://www.stsci.edu/hst/observatory/crds/calspec.html}
{http://www.stsci.edu/hst/observatory/crds/calspec.html}} to test the
performances of our bolometric corrections. This is becoming increasingly
relevant in the context of deriving reliable bolometric fluxes for
interferometric targets \citep[e.g.,][]{kara}.
Further, we use stellar parameters from the RAVE DR5 survey \citep{kunder} to
generate synthetic colours for $\sim 10^5$ stars, and compare them with the
observed ones, to validate our synthetic photometry at different wavelengths,
and over a wide range of stellar parameters. Star cluster fiducials in the
Pan-STARRS1 system are also used to show how our synthetic photometry can be
used together with theoretical stellar models to produce colour-magnitude
diagrams (CMDs).

\begin{figure}
\begin{center}
\includegraphics[width=0.45\textwidth]{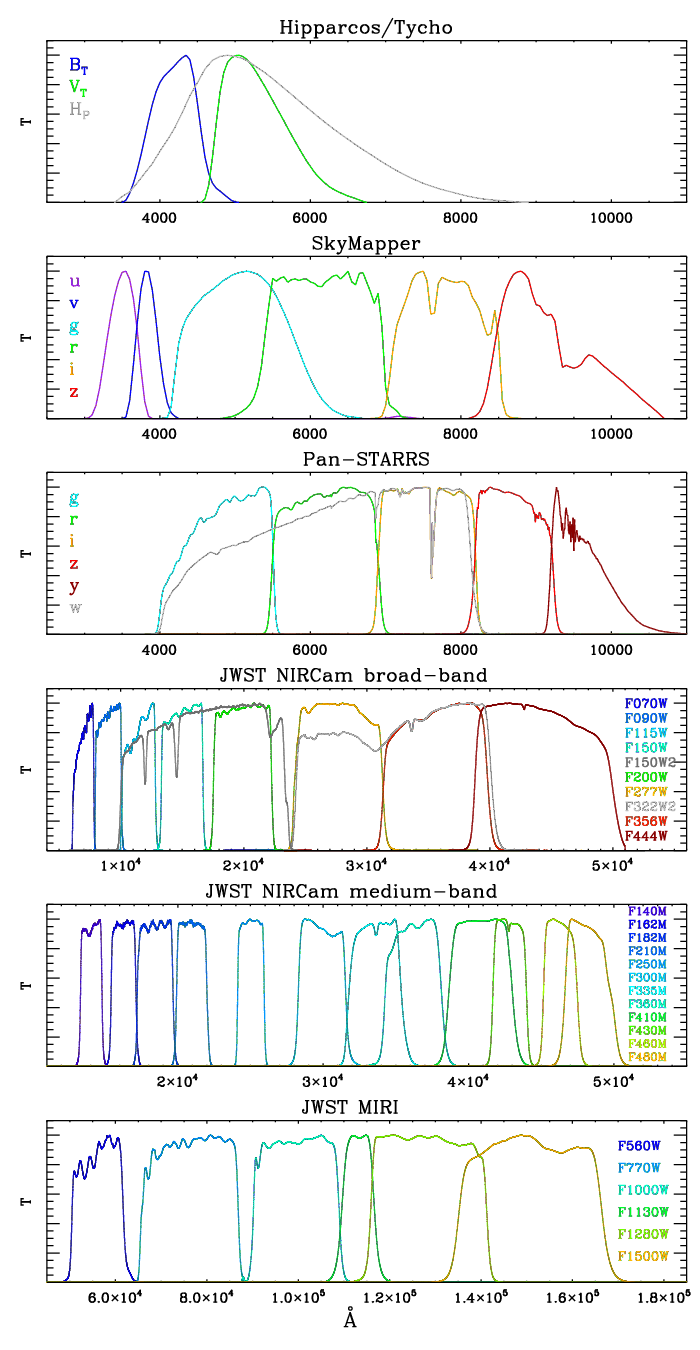}
\caption{System response functions from which synthetic colours and bolometric
  corrections have been computed. All curves are normalised to one, and shown
  as a function of wavelength (in \AA).}\label{f:filters} 
\end{center}
\end{figure}

\section{Synthetic photometry}\label{sec:sp}

The main concepts of synthetic photometry, along with the rationale behind the
{\tt Vega}, {\tt AB} and {\tt ST} magnitude systems are discussed in Paper I. 
Here we adhere to the same nomenclature, and provide information on the
filters transmission curves (Figure \ref{f:filters}), zero-points and absolute
calibrations used to compute synthetic colours and BCs
for the systems reported in Table \ref{t:t1} (analogous to that in Paper I). 

As in Paper I, we apply the \cite{ccm89} parametrisation of the extinction
law ($A_\lambda$) to synthetic spectra, in order to derive synthetic colours
and bolometric corrections for different values of reddening $E(B-V)$, between
0 and $0.72$. The
\cite{ccm89} law is described by a smooth power law in the range $0.3-3.3\mu$m,
and for the sake of this work we extrapolate its applicability to the reddest
{\it JWST}-NIRCam
filters, i.e. up to $\simeq 5\mu$m. This extrapolation is qualitatively
supported by recent observations, but beyond this limit the use of a single power
law breaks down, and the extinction law is characterised by emission features
\citep[e.g.][]{fritz}. Hence, it is not possible to apply a simple
parametrisation beyond this limit, but overall $A_\lambda$ is rather constant
when averaged over the wavelength range of a broad-band filter. Here, we assume
that the \cite{ccm89} law, which is parametrised as function of $x=1/\lambda$, 
remains constant at $A_{\lambda_x}$ longward of
$\lambda_x = \frac{1}{0.21} \mu$m. Because this assumption does not capture the
full complexity of the mid-infrared extinction, the effects of reddening given
in our tables for the MIRI system only provide some guidance of
how synthetic magnitudes and bolometric corrections will behave when $E(B-V)>0$.
{\it JWST}-MIRI filters centred on emission
features might have different magnitudes, and indeed, a full characterisation
of the extinction law in the mid-infrared will be one of the studies to be
carried out by the {\it JWST}. 

\begin{table*}
\centering
\caption{Characteristic parameters defining the photometric systems studied 
here. Refer to Paper I for the equations defining the {\tt VEGA}, {\tt AB}
and {\tt ST} system. The absolute calibration $\bar{f}_{\star,\zeta}$ is given
in $\ergs$. For each filter in {\it JWST}-NIRCam, the absolute calibration is
provided for module A (left), B (centre) and averaged AB (right). See text for
details.}\label{t:t1}
\begin{tabular}{c|l|ccc|c|c}
\hline 
        &            &       \multicolumn{3}{|c|}{{\tt VEGA} system}    & {\tt AB} system     & {\tt ST} system     \\
Filter  & $T_\zeta$   &            &                 &             &               &               \\
        &            & $m_{\star,\zeta}$ & $\bar{f}_{\star,\zeta}$      & $\epsilon_\zeta$  & $\epsilon_\zeta$    & $\epsilon_\zeta$    \\
\hline
 $H_P$     & 1      & $-$         & $-$                                       &     $-$     &     $+0.030$ &    $-$ \\
 $B_T$     & 1      & $-$         & $-$                                       &     $-$     &     $+0.100$ &    $-$ \\
 $V_T$     & 1      & $-$         & $-$                                       &     $-$     &     $+0.037$ &    $-$ \\
           &        &             &                                           &             &              &        \\
 $u$       & 2      & $-$         & $-$                                       &     $-$     &       $0$    &    $-$ \\
 $v$       & 2      & $-$         & $-$                                       &     $-$     &       $0$    &    $-$ \\
 $g$       & 2      & $-$         & $-$                                       &     $-$     &       $0$    &    $-$ \\
 $r$       & 2      & $-$         & $-$                                       &     $-$     &       $0$    &    $-$ \\
 $i$       & 2      & $-$         & $-$                                       &     $-$     &       $0$    &    $-$ \\
 $z$       & 2      & $-$         & $-$                                       &     $-$     &       $0$    &    $-$ \\
           &        &             &                                           &             &              &        \\
 $g$       & $3^\ast$      & $-$         & $-$                                       &     $-$     &       $0$    &    $-$ \\
 $r$       & $3^\ast$      & $-$         & $-$                                       &     $-$     &       $0$    &    $-$ \\
 $i$       & $3^\ast$      & $-$         & $-$                                       &     $-$     &       $0$    &    $-$ \\
 $z$       & $3^\ast$      & $-$         & $-$                                       &     $-$     &       $0$    &    $-$ \\
 $y$       & $3^\ast$      & $-$         & $-$                                       &     $-$     &       $0$    &    $-$ \\
 $w$       & $3^\ast$      & $-$         & $-$                                       &     $-$     &       $0$    &    $-$ \\
           &        &             &                                           &             &              &        \\
 $F070W$   & 4      & $0$         & 1.6709E-09 / 1.6785E-09 / 1.6747E-09      &     $0$     &       $0$    &    $0$ \\
 $F090W$   & 4      & $0$         & 8.2717E-10 / 8.3271E-10 / 8.2997E-10      &     $0$     &       $0$    &    $0$ \\
 $F115W$   & 4      & $0$         & 3.9736E-10 / 3.9663E-10 / 3.9700E-10      &     $0$     &       $0$    &    $0$ \\
 $F150W$   & 4      & $0$         & 1.5857E-10 / 1.5808E-10 / 1.5833E-10      &     $0$     &       $0$    &    $0$ \\
 $F150W2$  & 4      & $0$         & 1.2708E-10 / 1.2712E-10 / 1.2710E-10      &     $0$     &       $0$    &    $0$ \\
 $F200W$   & 4      & $0$         & 5.8441E-11 / 5.8466E-11 / 5.8454E-11      &     $0$     &       $0$    &    $0$ \\
 $F277W$   & 4      & $0$         & 1.7519E-11 / 1.6905E-11 / 1.7228E-11      &     $0$     &       $0$    &    $0$ \\
 $F322W2$  & 4      & $0$         & 1.0055E-11 / 1.0045E-11 / 1.0050E-11      &     $0$     &       $0$    &    $0$ \\
 $F356W$   & 4      & $0$         & 6.5505E-12 / 6.5012E-12 / 6.5257E-12      &     $0$     &       $0$    &    $0$ \\
 $F444W$   & 4      & $0$         & 2.9531E-12 / 2.8632E-12 / 2.9051E-12      &     $0$     &       $0$    &    $0$ \\
 $F140M$   & 4      & $0$         & 1.9808E-10 / 1.9869E-10 / 1.9839E-10      &     $0$     &       $0$    &    $0$ \\
 $F162M$   & 4      & $0$         & 1.1760E-10 / 1.1776E-10 / 1.1768E-10      &     $0$     &       $0$    &    $0$ \\
 $F182M$   & 4      & $0$         & 7.5682E-11 / 7.5715E-11 / 7.5699E-11      &     $0$     &       $0$    &    $0$ \\
 $F210M$   & 4      & $0$         & 4.7838E-11 / 4.7805E-11 / 4.7821E-11      &     $0$     &       $0$    &    $0$ \\
 $F250M$   & 4      & $0$         & 2.4566E-11 / 2.4601E-11 / 2.4583E-11      &     $0$     &       $0$    &    $0$ \\
 $F300M$   & 4      & $0$         & 1.2652E-11 / 1.2637E-11 / 1.2645E-11      &     $0$     &       $0$    &    $0$ \\
 $F335M$   & 4      & $0$         & 8.1407E-12 / 8.0504E-12 / 8.0975E-12      &     $0$     &       $0$    &    $0$ \\
 $F360M$   & 4      & $0$         & 6.0679E-12 / 6.0730E-12 / 6.0704E-12      &     $0$     &       $0$    &    $0$ \\
 $F410M$   & 4      & $0$         & 3.8715E-12 / 3.7956E-12 / 3.8333E-12      &     $0$     &       $0$    &    $0$ \\
 $F430M$   & 4      & $0$         & 3.1819E-12 / 3.1854E-12 / 3.1837E-12      &     $0$     &       $0$    &    $0$ \\
 $F460M$   & 4      & $0$         & 2.3336E-12 / 2.3538E-12 / 2.3441E-12      &     $0$     &       $0$    &    $0$ \\
 $F480M$   & 4      & $0$         & 2.0565E-12 / 1.9978E-12 / 2.0212E-12      &     $0$     &       $0$    &    $0$ \\
           &        &             &                                           &             &              &        \\
 $F560W$   & 5      & $0$         &              1.1113E-12                   &     $0$     &       $0$    &    $0$ \\
 $F770W$   & 5      & $0$         &              3.4150E-13                   &     $0$     &       $0$    &    $0$ \\
 $F1000W$  & 5      & $0$         &              1.1908E-13                   &     $0$     &       $0$    &    $0$ \\
 $F1130W$  & 5      & $0$         &              7.1065E-14                   &     $0$     &       $0$    &    $0$ \\
 $F1280W$  & 5      & $0$         &              4.3917E-14                   &     $0$     &       $0$    &    $0$ \\
 $F1500W$  & 5      & $0$         &              2.3114E-14                   &     $0$     &       $0$    &    $0$ \\
\hline
\end{tabular}
\begin{minipage}{1\textwidth}
An asterisk in the second column indicates that $T_\zeta$ given in the reference 
is already multiplied by $\lambda$ and renormalised (see discussion in Section 2.2 of Paper I).
--{\bf 1: Tycho}. Transmission curves from Bessell \& Murphy (2012);
$\epsilon_\zeta$ values are obtained  from their tables 3 and 5 (but with the
opposite sign, following our definition of $\epsilon_\zeta$ from Paper I).\,
--{\bf 2: SkyMapper}. Transmission curves from Bessell et al. (2011), under
the assumption that SkyMapper magnitudes are perfectly standardised on the
{\tt AB} system.\,
--{\bf 3: Pan-STARRS1}. Transmission curves from Tonry et al. (2012)
--{\bf 4: \emph{JWST}-NIRCam}. Transmission curves available at \href{https://jwst-docs.stsci.edu/display/JTI/NIRCam+Filters}{https://jwst-docs.stsci.edu/display/JTI/NIRCam+Filters}
--{\bf 5: \emph{JWST}-MIRI}. Transmission curves available at \href{http://ircamera.as.arizona.edu/MIRI/pces.htm}{http://ircamera.as.arizona.edu/MIRI/pces.htm}
\end{minipage}
\end{table*}

\subsection{Hipparcos/Tycho}

The Hipparcos catalogue contains photometry (and astrometry) for more than
$100,000$ stars measured with the $H_P$ band \citep{perryman97}. In addition,
for a much larger number of sources, a dichroic beam splitter on the Hipparcos
satellite sent light onto two photomultiplier tubes, providing simultaneously
measured $B_T$ and $V_T$ magnitudes. These are part of the Tycho-2 catalogue
\citep[{\it nota bene} the photometric system is called Tycho, but the
catalogue is Tycho-2 being the second and final data-release,][]{hog2000} which
contains photometry for 2.5 million stars. The Hipparcos and Tycho photometric
systems and passbands are discussed in \cite{vL97}, and have subsequently been
revised by \cite{bessell2000} and \cite{bm12}. Here, we generate synthetic
$H_P, B_T, V_T$
adopting the {\tt AB} formalism, with the passbands and zero-points provided by
\cite{bm12}. Some limitations of the observed $H_P$, $B_T$ and $V_T$ magnitudes
should be kept in mind. While these magnitudes are accurately and precisely
calibrated over the entire sky for the bulk of stars, the
quality of Tycho photometry quickly downgrades for $V_T \gtrsim 9$
\citep[e.g.,][their figure 5]{ruchti13}. In addition, $H_P$ photometry of the
brightest stars (say $H_P \lesssim 1.5$) is likely affected by pulse pileup and
should be used with caution \citep[][their figure 6]{bohlin14}.

\subsection{SkyMapper}

SkyMapper is a $1.35$m automated wide-field survey telescope located at Siding
Spring Observatory (Australia), to carry a multi-epoch photometric survey of the
entire southern sky to a final depth of $\simeq 20$ to $\simeq21$ mag
depending on the filters that are used. The SkyMapper
photometric system builds on the success of the $griz$ filters
used by the Sloan Digital Sky Survey \citep{fuku96}, with the added value of
the $uv$ bands, designed to be strongly sensitive to stellar parameters. The
SkyMapper $u$ band mimics the Str\"omgren $u$ filter, which is seated at the
Balmer discontinuity, and provides good temperature sensitivity in hot stars,
and gravity sensitivity across A, F and G spectral types. The SkyMapper $v$
band is similar to the DDO38 filter, which is metallicity sensitive. A
description and characterisation of the SkyMapper photometric system can be
found in \cite{skym}. The goal of SkyMapper is to map the entire southern sky
in all six filters. Currently, Data Release 1 (DR1) is available, which
covers more than $20,000\ \rm{deg}^2$ and has more than 300 million unique
sources. Further details on the survey, data quality and calibration are
provided in \cite{wolf}. It should be appreciated that the standardisation of
SkyMapper photometry to the {\tt AB} system is still in progress.  That is,
while our synthetic magnitudes have been computed assuming perfect adherence to
the {\tt AB} system, small differences may be present in observed data.
As a result, minor zero-point adjustments ($\epsilon_\zeta$) might be necessary
to match observations. Therefore, when comparing predicted colours with
published
photometry from a given Data Release, the accompanying paper should be checked
for information on potential zero-points departures from the {\tt AB} system. 

\subsection{Pan-STARRS1}

The Panoramic Survey Telescope \& Rapid Response System (Pan-STARRS1) employs a
$1.8$m telescope at Haleakala Observatories (Hawaii) to image three quarters
of the sky ($3\pi$ Survey) in the visible and near infrared broad-band filters
$grizy$ \citep{pan}. 
A thorough discussion of the the Pan-STARRS1 photometric system is given in
\cite{tonry}, who provide filter transmission curves, and detailed
information on how to achieve standardisation to the {\tt AB} magnitude system.

\subsection{\emph{JWST}}

The James Webb Space Telescope ({\it JWST}) will be the premier infrared space
observatory for the next
decade \citep{jwst}. The scientific instruments on board will
allow imaging, spectroscopy and coronography, covering a wide range of
wavelengths from the edge of the visible to the mid-infrared. The two imaging
instruments, the Near Infrared Camera \citep[NIRCam,][]{rieke05} and the Mid
Infrared Instrument \citep[MIRI,][]{rieke15} will deliver high-precision and
high-accuracy photometry across the $0.6-28\mu$m range. At the present time,
the system onto which {\it JWST} photometry will be standardised, and their
zero-points, is not yet clear.  Hence, in analogy to what is currently available
for the {\it HST} (see Paper I), we have generated magnitudes in the 
{\tt VEGA}, {\tt AB} and {\tt ST} systems also for {\it JWST} filters. The
zero-points of the {\tt AB} and {\tt ST} systems are known by definition.
(Standardising observations to the definition is the main observational
challenge.) To derive zero-points for the {\tt VEGA} system, we have used the
latest absolute flux for $\alpha$\,Lyr \citep{bohlin14} available from the
CALSPEC
library\footnote{Regularly updated absolute spectrophotometry can be found at
  http://www.stsci.edu/hst/observatory/cdbs/calspec.html. We have used the
  latest available at the time of this investigation,
  alpha\_lyr\_stis\_008.fits.}. While we expect that our synthetic photometry
provides a realistic representation of future {\it JWST} observations,
contaminations might still occur to the optical telescope elements during
deployment, thus changing the throughputs we have used here. Also, the
standardisation of real data should be checked to determine whether any
zero-point shifts should be applied onto our definitions. 

\subsubsection{NIRCam}\label{sec:nircam}

NIRCam is {\it JWST}'s primary imager in the wavelength range $0.6-5\mu$m. It
consists of two, nearly identical modules, A and B, and offers 29 bandpass
filters, covering extra-wide ($W2$), wide ($W$), medium ($M$) and narrow ($N$)
bands. A description of each filter and their science goals can be found at
\href{https://jwst-docs.stsci.edu/display/JTI/NIRCam+Filters}{https://jwst-docs.stsci.edu/display/JTI/NIRCam+Filters}.
We have generated colours and bolometric corrections for all but the
narrow filters, using the total system throughputs. Throughputs vary somewhat
between the 2 modules, especially at long wavelengths due to differences in
detector quantum efficiencies. Thus, we have computed synthetic NIRCam
photometry using module A, module B, and the average of the two (AB). The system
response functions plotted in Figure \ref{f:filters} are for the AB case.

\subsubsection{MIRI}

The imaging mode for {\it JWST}'s MIRI offers 9 broad-band filters from
$5.6-25.5\mu$m.
A description of each filter, and their purpose is available at
\href{https://jwst-docs.stsci.edu/display/JTI/MIRI+Imaging}{https://jwst-docs.stsci.edu/display/JTI/MIRI+Imaging}. Since MARCS synthetic fluxes extend only to
$20\mu$m, we have not computed synthetic photometry for filters redder than
$F1500W$ (Figure \ref{f:filters}). 

\begin{figure*}
\begin{center}
\includegraphics[width=0.9\textwidth]{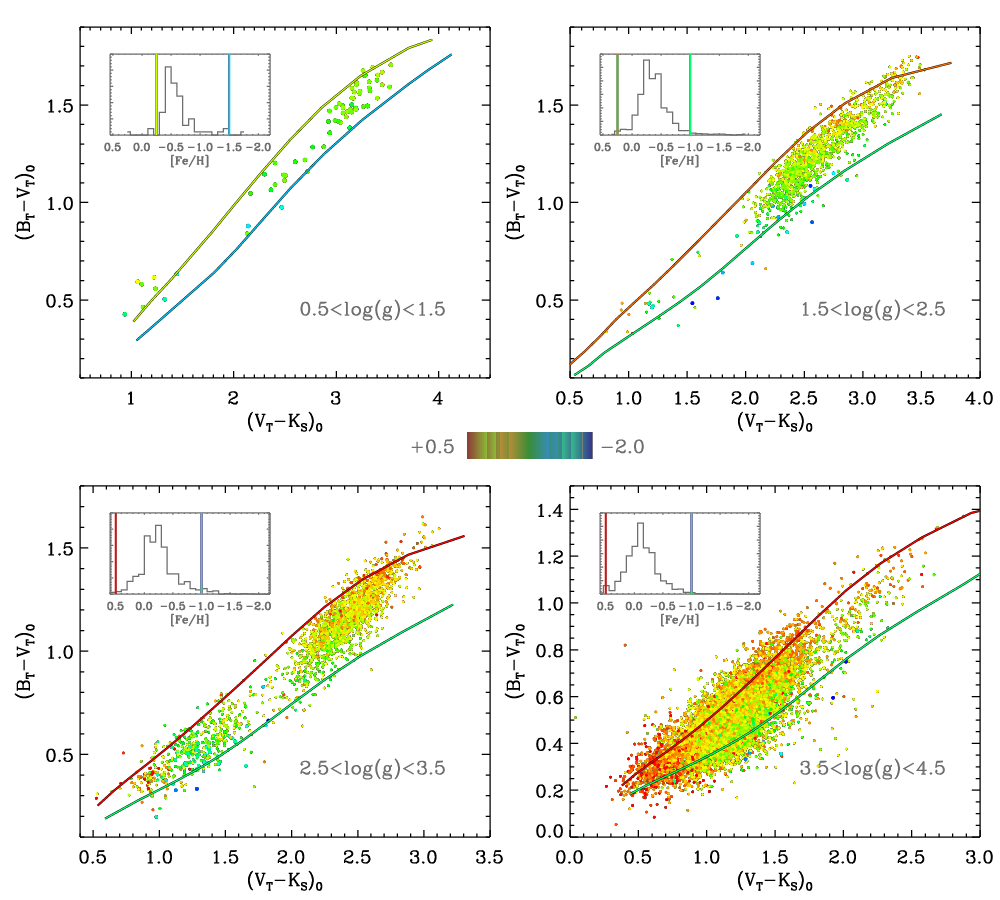}
\caption{Colour-colour planes for stars in RAVE DR5 satisfying the quality cuts
  described in the text, and also having Tycho and 2MASS photometry. Each
  panel displays stars in the $\logg$ range indicated in the bottom right,
  coded by metallicity as indicated by the palette. Insets show the
  metallicity histogram within each panel. Continuous lines are synthetic
  colours predicted by MARCS models at metallicities broadly encompassing
  the sample (marked by vertical lines in histograms). $\logg$ of MARCS models
  is $1, 2, 4, 3$ (clockwise from top-left). Only stars with photometric
  errors $<0.05$~mag in each band are shown. Photometry has been dereddened
  using rescaled $E(B-V)$ values as described in the text.}\label{f:col4}
\end{center}
\end{figure*}

\begin{figure*}
\begin{center}
\includegraphics[width=0.9\textwidth]{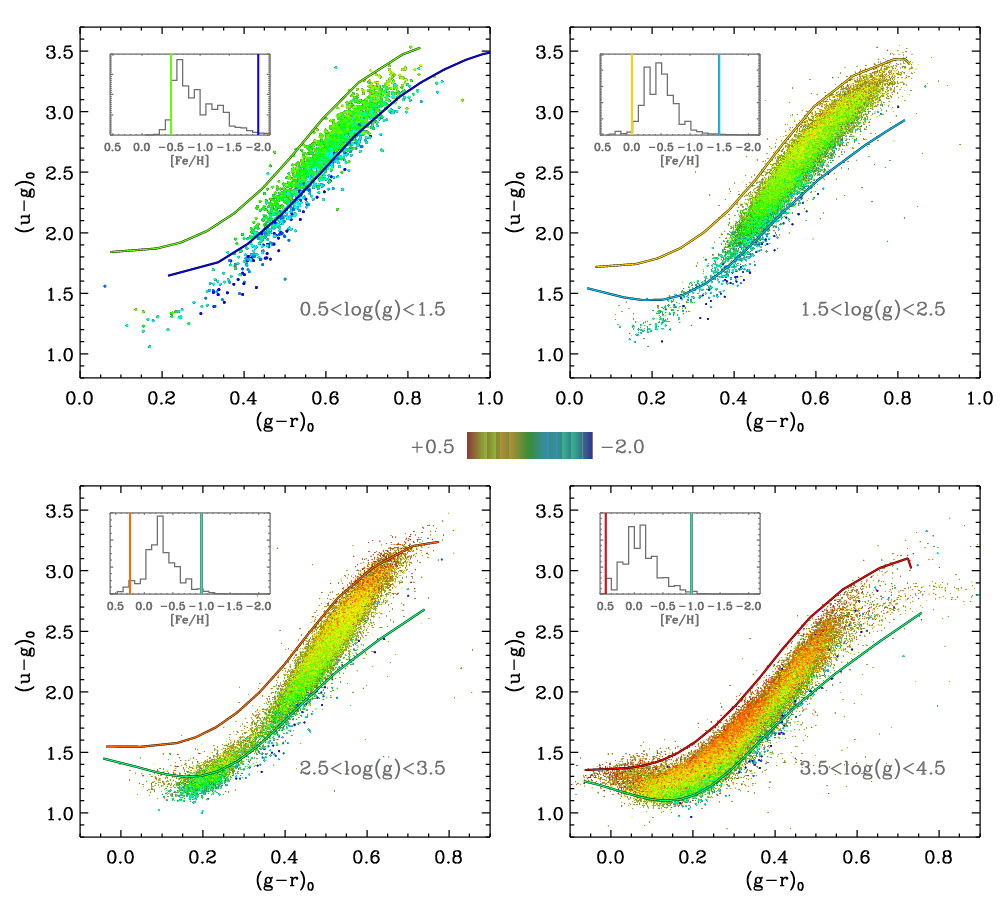}
\caption{Same as Figure \ref{f:col4} but for combinations of SkyMapper colours.
  In addition to the RAVE quality cuts described in the main text, only stars
  with photometric errors $<0.05$~mag in each band, no photometric flags, and
  {\tt class\_star$>0.9$} are shown. Photometry has been dereddened using
  rescaled $E(B-V)$.}\label{f:col1}
\end{center}
\end{figure*}

\section{Conversion to physical fluxes}\label{sec:phy}

In certain situations, it might be useful to transform a bolometric
correction into a bolometric flux. For example, when the apparent magnitude of
a star of is known, and its angular diameter is measured, then knowledge of
the bolometric flux allows a direct determination of its $\teff$
\citep[e.g,][]{hanbury67}. Alternatively, one might wish to know the
bolometric flux of a star, to
compare with luminosities predicted from stellar models if the distance is
known. From the definition of bolometric correction in a given band $\zeta$:
\begin{equation}\label{eq:BCdef}
BC_\zeta = m_{\rm{Bol}}-m_\zeta = M_{\rm{Bol}}-M_\zeta
\end{equation}
which is the same whether
applied to apparent (lower case) or absolute (upper case) magnitudes. However,
the zero-point of both apparent and absolute bolometric magnitudes is defined
by the solar absolute bolometric magnitude ($M_{\rm{Bol},\odot}$) and luminosity
($L_\odot$):
\begin{equation}
M_{\rm{Bol}} = -2.5 \log\frac{L}{L_\odot} + M_{\rm{Bol},\odot}.
\end{equation}
We stress that while the solar luminosity $L_\odot$ is a measured
quantity,
$M_{\rm{Bol},\odot}$ is an arbitrary zero-point and any value is equally
legitimate on the condition that once chosen, all bolometric corrections are
scaled accordingly. To keep consistency with Paper I, our tables of bolometric
corrections and interpolation routines are computed fixing
$M_{\rm{Bol},\odot}=4.75$, and this value must be used in
Eq.~(\ref{eq:BCdef})--(\ref{eq:bc}).
It can be shown that the bolometric flux
of a star having an observed magnitude $m_\zeta$ and bolometric correction
$BC_\zeta$ is then:
\begin{displaymath}
f_{\rm{Bol}}\,(\flux) =\frac{\pi L_\odot}{(1.296\times 10^9\,\rm{AU})^2} 10^{-0.4(BC_\zeta-M_{\rm{Bol},\odot}+m_\zeta-10)}
\end{displaymath}
\begin{equation}\label{eq:bc}
\simeq 8.358\times 10^{-45} L_\odot\,10^{-0.4(BC_\zeta-M_{\rm{Bol},\odot}+m_\zeta-10)},
\end{equation}
where we adopted $\rm{AU}=1.495978707 \times 10^{13}$ cm from the IAU 2012
Resolution B2, and $L_\odot$ is the solar luminosity in $\rm{erg\,s^{-1}}$.
Note that the solar luminosity specified in IAU 2015 Resolution B3 is reduced
by $\simeq 0.4$ percent with respect to older measurements\footnote{The relevant
IAU Resolutions can be found at \href{https://www.iau.org/static/resolutions/IAU2012_English.pdf}{https://www.iau.org/static/resolutions/IAU2012\_English.pdf} \href{https://www.iau.org/static/resolutions/IAU2015_English.pdf}{https://www.iau.org/static/resolutions/IAU2015\_English.pdf}.\\ These
resolutions also recommend the adoption of $M_{\rm{Bol},\odot}=4.74$. If users wish to adopt this value, the BCs from our interpolation routines must be increased by $-0.01$\ mag}.
Also worth mentioning is that, when deriving bolometric fluxes to compare with
the predictions of a given set of stellar models, the same value of
$L_\odot$ that was assumed in the computation of those models should be used in
Eq.~(\ref{eq:bc}) in order to obtain internal consistency. Finally, we recall
that if a value of $E(B-V)$ is provided, our interpolation routines take into
account the effect of reddening on bolometric corrections, which can then be
directly used in Eq.~(\ref{eq:bc}). Alternatively, one could use bolometric
corrections for no reddening, and deredden apparent magnitudes as
$m_\zeta-R_\zeta E(B-V)$, using the extinction coefficients $R_\zeta$ tabulated in
Table A1 of Paper I, and Table \ref{tab:tabler} in the Appendix (the first
method is to be preferred because it automatically takes into account the
dependence of extinction coefficients on stellar parameters, whereas using
tabulated $R_\zeta$ is not as exact. See discussion in Paper I). 

It might also be useful to deal with physical units in the case of infrared
observations, where Jansky is often the adopted unit of measurement instead of
magnitude. This might be relevant for the {\it JWST} filters, and the
fact that we provide {\tt AB} magnitudes readily allows for this conversion.
From the definition of {\tt AB} magnitudes, it follows that:
\begin{equation}
\bar{f_\nu}({\rm Jy})= 3631 \times 10^{-0.4(m_{\tt AB}-\epsilon_\zeta)}.
\end{equation}
where $m_{\tt AB}$ is the {\tt AB} magnitude for a given $\zeta$ filter and
$\epsilon_\zeta$ allows for zero-point offsets (with $\epsilon_\zeta=0$ for a
perfect standardisation to the {\tt AB} system, see the discussion in Paper I).

\section{Comparisons with observations}

Here we evaluate the performance of our BCs and synthetic colours in three
different ways. The first method relies on having precise and accurate stellar
parameters, observed magnitudes and absolute spectrophotometry (which is
currently possible only for a limited number of stars) to check how well
bolometric fluxes can be recovered.  The second
depends instead on having a statistically large sample of stars
(of order $10^5$) with overall well calibrated stellar parameters. In this case,
comparisons of synthetic and observed colours in different bands for a wide
range of $\teff$, $\logg$ and $\feh$ test the combined performance of the MARCS
model fluxes together with the standardisation, zero-points and system response
functions that were adopted when we computed the synthetic photometry.
In the third approach, we couple our BCs with stellar isochrones to examine
the extent to which the stellar models are able to provide consistent
interpretations of as many CMDs as possible that can be constructed from
multi-colour observations of a given star cluster.  Such analyses can provide
valuable insights into both the temperatures of the stellar models and
possible deficiencies of the synthetic colours. In all instances we have used
the MARCS tables of BCs and interpolation routines made publicly available
through this work, assuming standard $\aFe$ enhancements. As for Paper I, BC
transformations for different $\alpha$-enhancements can also be generated (see
also Appendix A).

\begin{figure*}
\begin{center}
\includegraphics[width=0.9\textwidth]{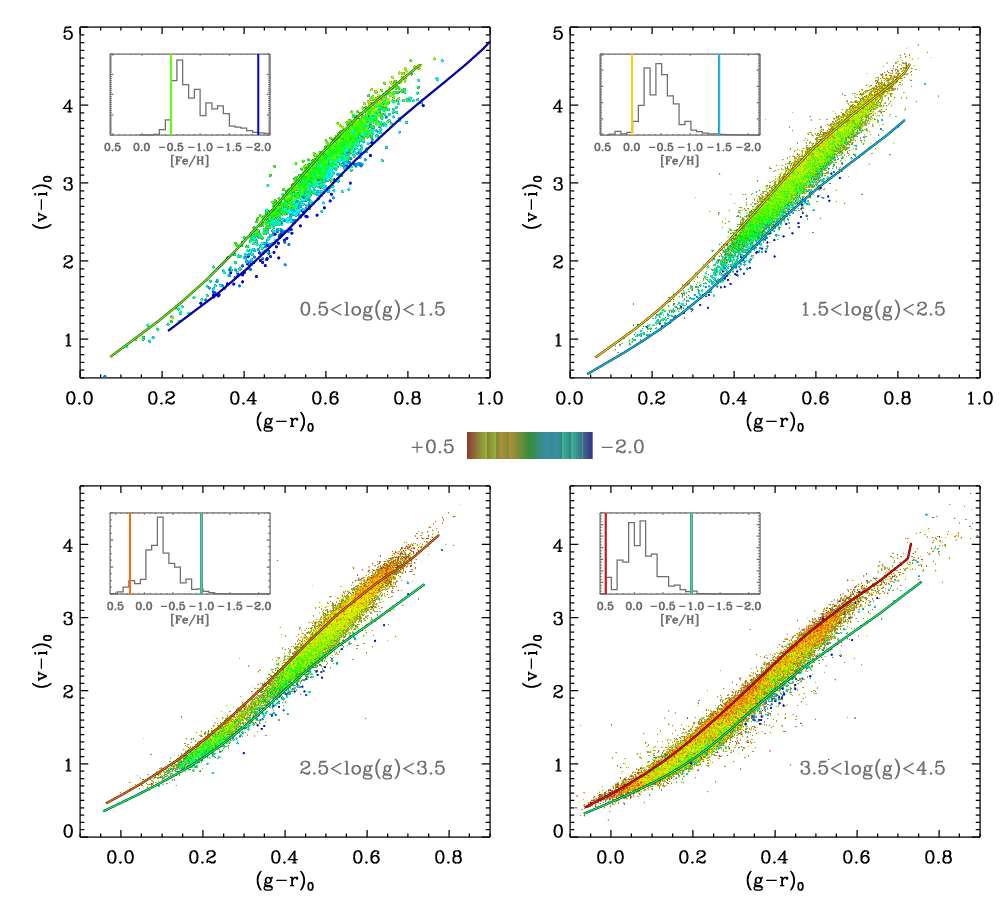}
\caption{Same as Figure \ref{f:col1} but for a different combination of
  SkyMapper colours.}\label{f:col2}
\end{center}
\end{figure*}

\begin{figure*}
\begin{center}
\includegraphics[width=0.9\textwidth]{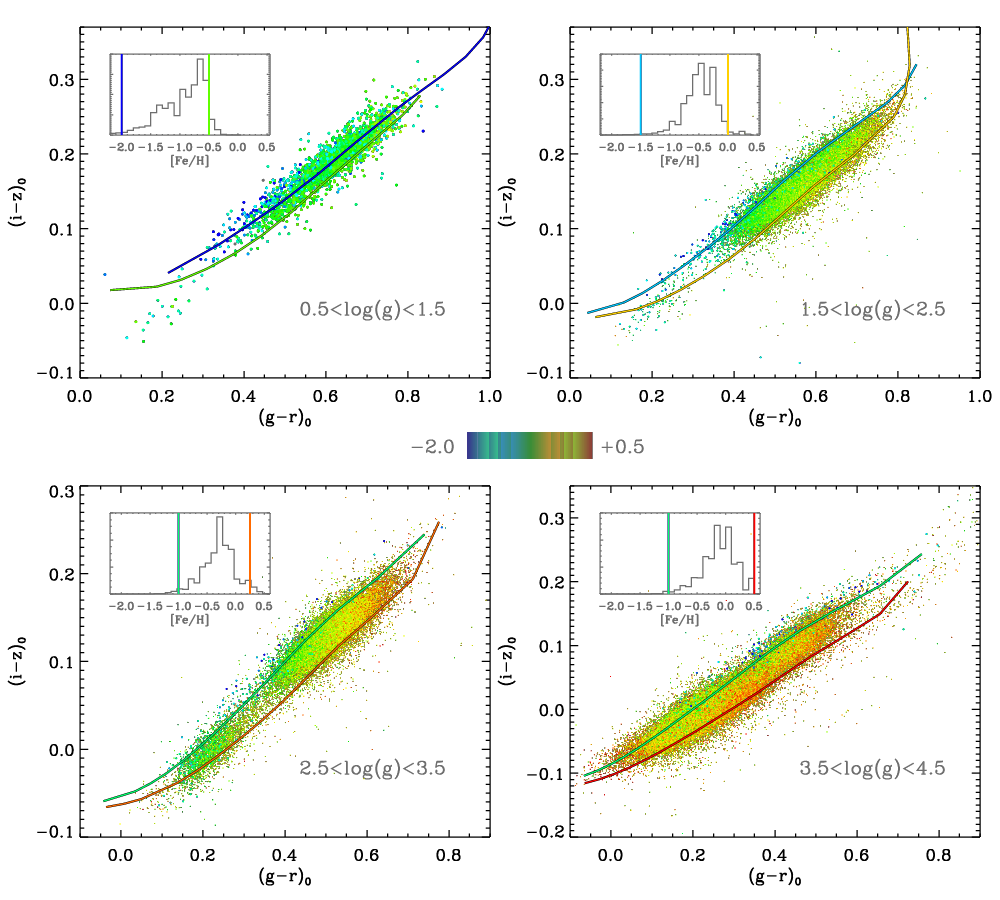}
\caption{Same as Figure \ref{f:col1} but for a different combination of
  SkyMapper colours.}\label{f:col3}
\end{center}
\end{figure*}

\subsection{Testing the bolometric flux scale}

The CALSPEC library contains composite stellar spectra that are flux
standards on the {\it HST} system. The latter is based on three hot, pure
hydrogen
white dwarf standards normalised to the absolute flux of Vega at 5556 \AA.
This spectrophotometry is expected to be accurate at the (few) percent level
\citep{bohlin14}. 

Thus, CALSPEC absolute spectrophotometry currently provides the best way to
test how well our bolometric corrections can recover stellar fluxes. In
particular, the highest quality measurements in the CALSPEC library are those
obtained by the STIS ($0.17-1.01\mu$m) and NICMOS ($1.01-2.49\mu$m) instruments
on board of the {\it HST}.

Since our synthetic BCs are computed at a given $\teff$,
$\logg$, $\feh$ and $\aFe$, these quantities need to be known. Also,
we want to consider stars with good photometry, no binary
contamination, no dust emission, no reported variability and ideally little or
no reddening. These requirements are met at different levels by a number of
stars in the CALSPEC library, and for this reason we group them into three
categories. A summary of all stars, their adopted parameters, CALSPEC
bolometric fluxes and those we recover using our BCs is presented in
Table \ref{t:t2}. In what follows we briefly discuss each category.

\subsubsection{Tier 1}

Tier 1 stars are those with best possible parameters. In fact, only
one star, HD209458, qualifies. It has both STIS and NICMOS
spectrophotometry, meaning that almost 96 percent of its bolometric flux is
directly measured. This star is a primary CALSPEC calibrator, and a proposed
JWST primary flux standard \citep{gordon}.
Its stellar parameters are extremely well characterised,
which then lead to a reliable estimate of its bolometric flux \citep{dBA}.
Also, this star is nearby ($\simeq 50$~pc) and hence reddening free, and it
has very good photometry in the Tycho \citep{hog2000} and 2MASS \citep{scs06}
systems.

Using the BCs that are obtained by interpolations in our tables, we are usually
able to recover the CALSPEC bolometric flux of this star to within $\simeq 1$
percent in each band. To account for the uncertainties associated with
the bolometric fluxes so derived, we ran MonteCarlo simulations that take
into account the quoted uncertainties in both the input stellar parameters and
the observed photometry. This source of uncertainty amounts to of order 2 percent
in each band. 

\subsubsection{Tier 2}

Stars in this category still have good stellar parameters, but they lack the
full optical and infrared spectrophotometric coverage. They all have STIS
observations between $1700$ and $10000$~\AA, a region which encompasses
about 70 percent of the bolometric flux, as well as most of the absorption
lines in these stars.
Longward of $10000$~\AA\, the CALSPEC spectrophotometry has been extended
with tailored synthetic fluxes \citep{bohlin14,bohlin17}. All stars in this
category are nearby ($\sim$ 25-75~pc), and hence located within the local
bubble where reddening is virtually non-existent
\citep[e.g.,][]{leroy93,lallement03}. The absence of
any measurable reddening is confirmed by $H\beta$ photometry
\citep{holmberg07}. For all of these stars we adopt stellar parameters from the
latest revision of the Geneva-Copenhagen Survey \citep{c11}.

For stars in this category as well, we usually recover their CALSPEC bolometric
fluxes to within $\simeq 1$ percent in each band, although occasionally the
agreement downgrades to approximately
5 percent. Flux uncertainties estimated from MonteCarlo simulations are usually
larger that the differences with respect to CALSPEC, suggesting that the
precision at which we recover bolometric fluxes is higher than that inferred
from the MonteCarlo analyses. 

\begin{table*}
\centering
\caption{CALSPEC bolometric fluxes ($\flux$) vs. those recovered interpolating
  our tables at the given $\teff$, $\logg$ and $\feh$ in a number
  of filters. In Eq.(\ref{eq:bc}) we have adopted $L_\odot = 3.842 \times 10^{33} \rm{erg\,s^{-1}}$
  \citep{bahcall06} which at $R_\odot=6.957 \times 10^{10}$~cm yields $T_{\rm{eff},\odot}=5777$~K consistent with
  the solar effective temperature in the MARCS library.}\label{t:t2}
\begin{tabular}{cccccccrrr}
\hline 
 Star    & Tier & $\teff$~(K) &  $\logg$  &  $\feh$  &  $E(B-V)$  & $\fbol$ \tiny{CALSPEC} &  $\fbol$ \tiny{MARCS} & $\sigma(\%)$ & $\Delta(\%)$ \\
\hline
HD209458 & 1 &  $6070$  &   $4.38$  &  $0.00$  &     0      &  $2.289 \times 10^{-8}$ &  $<> : 2.285 \times 10^{-8}$  & &   $-0.18\%$      \\
         &   &          &           &          &            &                        &   $H_P: 2.259 \times 10^{-8}$  & $1.38$ &   $-1.32\%$     \\
         &   &          &           &          &            &                        &   $B_T: 2.257 \times 10^{-8}$  & $1.94$ &   $-1.42\%$     \\
         &   &          &           &          &            &                        &   $V_T: 2.296 \times 10^{-8}$  & $1.24$ &   $0.32\%$      \\
         &   &          &           &          &            &                        &   $J: 2.330 \times 10^{-8}$    & $2.03$ &   $1.81\%$      \\
         &   &          &           &          &            &                        &   $H: 2.303 \times 10^{-8}$    & $3.67$ &   $0.60\%$      \\
         &   &          &           &          &            &                        &   $K_S: 2.265 \times 10^{-8}$  & $2.66$ &   $-1.05\%$     \\
\hline
HD31128  & 2 &  $6093$  &   $4.46$  &  $-1.58$ &     0      &  $6.429 \times 10^{-9}$  &  $<> : 6.427 \times 10^{-9}$ & &   $-0.02\%$      \\
         &   &        &           &          &            &                         &   $H_P: 6.354 \times 10^{-9}$  & $1.07$ &   $-1.16\%$     \\
         &   &        &           &          &            &                         &   $B_T: 6.568 \times 10^{-9}$  & $4.20$ &   $2.17\%$     \\
         &   &        &           &          &            &                         &   $V_T: 6.395 \times 10^{-9}$  & $1.80$ &   $-0.52\%$     \\
         &   &        &           &          &            &                         &   $J: 6.653 \times 10^{-9}$    & $4.85$ &   $3.50\%$     \\
         &   &        &           &          &            &                         &   $H: 6.389 \times 10^{-9}$    & $6.58$ &   $-0.61\%$     \\
         &   &        &           &          &            &                         &   $K_S: 6.360 \times 10^{-9}$  & $6.68$ &   $-1.07\%$     \\
         &   &        &           &          &            &                         &   $B: 6.227 \times 10^{-9}$    & $3.70$ &   $-3.14\%$     \\
         &   &        &           &          &            &                         &   $V: 6.425 \times 10^{-9}$    & $1.60$ &   $-0.06\%$     \\
         &   &        &           &          &            &                         &   $R_C: 6.419 \times 10^{-9}$  & $1.72$ &   $-0.15\%$     \\
         &   &        &           &          &            &                         &   $I_C: 6.484 \times 10^{-9}$  & $2.64$ &   $0.86\%$     \\
\hline
HD106252 &  2 & $5903$  &   $4.38$  &  $-0.06$ &     0      &  $2.903 \times 10^{-8}$  &  $<> : 2.896  \times 10^{-8}$ & &   $-0.25\%$   \\
         &    &       &           &          &            &                         &   $H_P: 2.845 \times 10^{-8}$   & $0.92$ &   $-1.98\%$    \\
         &    &       &           &          &            &                         &   $B_T: 2.843 \times 10^{-8}$   & $3.29$ &   $-2.05\%$    \\
         &    &       &           &          &            &                         &   $V_T: 2.878 \times 10^{-8}$   & $1.38$ &   $-0.87\%$    \\
         &    &       &           &          &            &                         &   $J  : 2.894 \times 10^{-8}$   & $2.93$ &   $-0.32\%$    \\
         &    &       &           &          &            &                         &   $H  : 2.947 \times 10^{-8}$   & $5.14$ &   $1.54\%$     \\
         &    &       &           &          &            &                         &   $K_S: 2.967 \times 10^{-8}$   & $3.79$ &   $2.19\%$     \\
\hline
HD159222 & 2 &  $5786$  &   $4.37$  &  $0.15$ &      0      &  $6.568 \times 10^{-8}$  &  $<> : 6.574  \times 10^{-8}$ & &  $0.09\%$     \\
         &   &        &           &          &            &                         &   $H_P: 6.518  \times 10^{-8}$  & $0.89$ &   $-0.76\%$    \\
         &   &        &           &          &            &                         &   $B_T: 6.666 \times 10^{-8}$   & $3.07$ &   $1.49\%$     \\
         &   &        &           &          &            &                         &   $V_T: 6.550 \times 10^{-8}$   & $1.24$ &   $-0.27\%$     \\
         &   &        &           &          &            &                         &   $J:   6.684 \times 10^{-8}$   & $2.50$ &   $1.77\%$     \\
         &   &        &           &          &            &                         &   $H:   6.629 \times 10^{-8}$   & $3.35$ &   $0.93\%$     \\
         &   &        &           &          &            &                         &   $K_S: 6.568 \times 10^{-8}$   & $2.83$ &   $0.00\%$     \\
         &   &        &           &          &            &                         &   $B:   6.532 \times 10^{-8}$   & $2.80$ &   $-0.55\%$    \\
         &   &        &           &          &            &                         &   $V:   6.442 \times 10^{-8}$   & $1.62$ &   $-1.91\%$    \\
\hline
HD185975 & 2 & $5671$  &   $4.01$  &  $0.06$ &      0      &  $1.587 \times 10^{-8}$  &  $<> : 1.574 \times 10^{-8}$  & &   $-0.77\%$     \\
         &   &        &           &          &            &                         &   $H_P: 1.551 \times 10^{-8}$  & $1.26$ &   $-2.26\%$    \\
         &   &        &           &          &            &                         &   $B_T: 1.564 \times 10^{-8}$  & $4.11$ &   $-1.43\%$    \\
         &   &        &           &          &            &                         &   $V_T: 1.553 \times 10^{-8}$  & $1.62$ &   $-2.16\%$    \\
         &   &        &           &          &            &                         &   $J:   1.598 \times 10^{-8}$  & $2.83$ &   $0.68\%$     \\
         &   &        &           &          &            &                         &   $H:   1.586 \times 10^{-8}$  & $3.77$ &   $-0.06\%$    \\
         &   &        &           &          &            &                         &   $K_S: 1.596 \times 10^{-8}$  & $4.16$ &   $0.58\%$     \\
\hline
HD205905 & 2 & $5991$  &   $4.49$  &  $0.09$  &      0     &  $5.374 \times 10^{-8}$  &  $<> :  5.249 \times 10^{-8}$ & &   $-2.32\%$     \\
         &   &        &           &          &            &                         &   $H_P: 5.179 \times 10^{-8}$  & $1.34$ &   $-3.64\%$    \\
         &   &        &           &          &            &                         &   $B_T: 5.075 \times 10^{-8}$  & $4.75$ &   $-5.57\%$    \\
         &   &        &           &          &            &                         &   $V_T: 5.222 \times 10^{-8}$  & $1.60$ &   $-2.83\%$    \\
         &   &        &           &          &            &                         &   $J:   5.165 \times 10^{-8}$  & $3.68$ &   $-3.90\%$    \\
         &   &        &           &          &            &                         &   $H:   5.463 \times 10^{-8}$  & $5.32$ &   $1.65\%$     \\
         &   &        &           &          &            &                         &   $K_S: 5.393 \times 10^{-8}$  & $5.46$ &   $0.35\%$     \\
\hline
HD37962  & 2 & $5756$  &   $4.46$  & $-0.26$  &      0     &  $2.010 \times 10^{-8}$  &  $<> :  1.987 \times 10^{-8}$  & &   $-1.13\%$     \\
         &   &        &           &          &            &                         &   $H_P:  1.974 \times 10^{-8}$  & $1.19$ &   $-1.83\%$    \\
         &   &        &           &          &            &                         &   $B_T:  1.921 \times 10^{-8}$  & $3.94$ &   $-4.46\%$    \\
         &   &        &           &          &            &                         &   $V_T:  1.980 \times 10^{-8}$  & $1.59$ &   $-1.51\%$    \\
         &   &        &           &          &            &                         &   $J:    1.980 \times 10^{-8}$  & $2.93$ &   $-1.51\%$    \\
         &   &        &           &          &            &                         &   $H:    2.041 \times 10^{-8}$  & $4.59$ &   $1.53\%$     \\
         &   &        &           &          &            &                         &   $K_S:  2.030 \times 10^{-8}$  & $4.16$ &   $0.97\%$     \\
\hline
\end{tabular}
\end{table*}
\setcounter{table}{1}
\begin{table*}
\centering
\caption{Continued}
\begin{tabular}{cccccccrrr}
\hline 
 Star    & Tier & $\teff$~(K) &  $\logg$  &  $\feh$  &  $E(B-V)$  & $\fbol$ \tiny{CALSPEC} &  $\fbol$ \tiny{MARCS} & $\sigma(\%)$ &  $\Delta(\%)$ \\
\hline
HD38949   & 2 & $6065$  &   $4.47$  &  $-0.16$  &     0      &  $2.035 \times 10^{-8}$ &  $<> : 2.002 \times 10^{-8}$  & &   $-1.66\%$      \\
          &   &        &           &          &            &                        &   $H_P: 1.981 \times 10^{-8}$  & $1.13$ &   $-2.69\%$     \\
          &   &        &           &          &            &                        &   $B_T: 1.938 \times 10^{-8}$  & $3.94$ &   $-4.76\%$     \\
          &   &        &           &          &            &                        &   $V_T: 1.985 \times 10^{-8}$  & $1.46$ &   $-2.46\%$     \\
          &   &        &           &          &            &                        &   $J:   2.030 \times 10^{-8}$  & $3.81$ &   $-0.28\%$     \\
          &   &        &           &          &            &                        &   $H:   2.069 \times 10^{-8}$  & $5.26$ &   $ 1.67\%$     \\
          &   &        &           &          &            &                        &   $K_S: 2.006 \times 10^{-8}$  & $4.84$ &   $-1.46\%$     \\
\hline
BD$+02$ 3375 & 3 & $6163$   & $4.13$    &  $-2.24$ &  $0.034$   &  $3.446\times 10^{-9}$  &  $<> : 3.386 \times 10^{-9}$  & &   $-1.75\%$     \\
             &   &        &           &          &            &                        &   $H_P: 3.385 \times 10^{-9}$  & $0.90$ &   $-1.76\%$     \\
             &   &        &           &          &            &                        &   $B_T: 3.369 \times 10^{-9}$  & $3.71$ &   $-2.25\%$     \\
             &   &        &           &          &            &                        &   $V_T: 3.250 \times 10^{-9}$  & $2.64$ &   $-5.69\%$     \\
             &   &        &           &          &            &                        &   $J: 3.498 \times 10^{-9}$    & $3.66$ &   $1.52\%$      \\
             &   &        &           &          &            &                        &   $H: 3.431 \times 10^{-9}$    & $6.03$ &   $-0.43\%$     \\
             &   &        &           &          &            &                        &   $K_S: 3.328 \times 10^{-9}$  & $5.86$ &   $-3.41\%$     \\
             &   &        &           &          &            &                        &   $B: 3.372 \times 10^{-9}$    & $2.96$ &   $-2.16\%$     \\
             &   &        &           &          &            &                        &   $V: 3.419 \times 10^{-9}$    & $1.54$ &   $-0.79\%$     \\
             &   &        &           &          &            &                        &   $R_C: 3.393 \times 10^{-9}$  & $1.49$ &   $-1.52\%$     \\
             &   &        &           &          &            &                        &   $I_C: 3.412 \times 10^{-9}$  & $2.05$ &   $-0.98\%$     \\
\hline
BD$+21$ 0607 & 3 & $6285$   & $4.29$    &  $-1.56$ &  $0.020$   &  $6.228 \times 10^{-9}$  &  $<> : 6.249 \times 10^{-9}$  & &   $0.34\%$     \\
             &   &        &           &          &            &                        &   $H_P: 6.145 \times 10^{-9}$  & $0.88$ &    $-1.33\%$    \\
             &   &        &           &          &            &                        &   $B_T: 6.514 \times 10^{-9}$  & $3.27$ &    $4.59\%$    \\
             &   &        &           &          &            &                        &   $V_T: 6.007 \times 10^{-9}$  & $1.96$ &    $-3.55\%$    \\
             &   &        &           &          &            &                        &   $J:   6.354 \times 10^{-9}$  & $3.53$ &    $2.02\%$    \\
             &   &        &           &          &            &                        &   $H:   6.261 \times 10^{-9}$  & $4.91$ &    $0.53\%$    \\
             &   &        &           &          &            &                        &   $K_S: 6.204 \times 10^{-9}$  & $5.07$ &    $-0.39\%$    \\
             &   &        &           &          &            &                        &   $B:   6.244 \times 10^{-9}$  & $2.88$ &    $0.25\%$    \\
             &   &        &           &          &            &                        &   $V:   6.290 \times 10^{-9}$  & $1.56$ &    $0.99\%$    \\
             &   &        &           &          &            &                        &   $R_C: 6.198 \times 10^{-9}$  & $1.59$ &    $-0.48\%$    \\
             &   &        &           &          &            &                        &   $I_C: 6.273 \times 10^{-9}$  & $2.13$ &    $0.72\%$    \\
\hline
BD$+29$ 2091 & 3 & $5974$   & $4.58$    &  $-1.99$ &  $0.004$   &  $2.363 \times 10^{-9}$  &  $<> : 2.354 \times 10^{-9}$  & &    $-0.36\%$     \\
             &   &        &           &          &            &                        &   $H_P: 2.324 \times 10^{-9}$  & $0.81$ &    $-1.62\%$    \\
             &   &        &           &          &            &                        &   $B_T: 2.294 \times 10^{-9}$  & $5.33$ &    $-2.90\%$    \\
             &   &        &           &          &            &                        &   $V_T: 2.226 \times 10^{-9}$  & $4.22$ &    $-5.81\%$    \\
             &   &        &           &          &            &                        &   $J:   2.416 \times 10^{-9}$  & $3.66$ &    $2.24\%$    \\
             &   &        &           &          &            &                        &   $H:   2.420 \times 10^{-9}$  & $4.66$ &    $2.43\%$    \\
             &   &        &           &          &            &                        &   $K_S: 2.405 \times 10^{-9}$  & $5.12$ &    $1.77\%$    \\
             &   &        &           &          &            &                        &   $B:   2.350 \times 10^{-9}$  & $3.03$ &    $-0.54\%$    \\
             &   &        &           &          &            &                        &   $V:   2.400 \times 10^{-9}$  & $1.50$ &    $1.58\%$    \\

\hline
BD$+54$ 1216 & 3 & $6127$   & $4.29$    &  $-1.63$ &  $0.002$   &  $ 3.812 \times 10^{-9}$  &  $<> :  3.816 \times 10^{-9}$ & &    $0.09\%$     \\
             &   &        &           &          &            &                        &   $H_P:  3.775 \times 10^{-9}$    & $0.89$ &    $-0.98\%$    \\
             &   &        &           &          &            &                        &   $B_T:  3.773 \times 10^{-9}$    & $3.61$ &    $-1.04\%$    \\
             &   &        &           &          &            &                        &   $V_T:  3.994 \times 10^{-9}$    & $2.29$ &    $4.78\%$    \\
             &   &        &           &          &            &                        &   $J:  3.936 \times 10^{-9}$      & $3.73$ &     $3.24\%$    \\
             &   &        &           &          &            &                        &   $H:  3.790 \times 10^{-9}$      & $4.76$ &    $-0.58\%$    \\
             &   &        &           &          &            &                        &   $K_S:  3.759 \times 10^{-9}$    & $4.94$ &    $-1.40\%$    \\
             &   &        &           &          &            &                        &   $B:  3.748 \times 10^{-9}$      & $2.98$ &    $-1.67\%$    \\
             &   &        &           &          &            &                        &   $V:  3.801 \times 10^{-9}$      & $1.55$ &    $-0.31\%$    \\
             &   &        &           &          &            &                        &   $R_C:  3.797 \times 10^{-9}$    & $1.59$ &    $-0.40\%$    \\
             &   &        &           &          &            &                        &   $I_C:  3.783 \times 10^{-9}$    & $2.14$ &    $-0.76\%$    \\
\hline
HD74000      & 3 & $6362$   & $4.12$    &  $-2.01$ &  $0.003$   &  $ 4.005 \times 10^{-9}$  &  $<> :  3.965 \times 10^{-9}$ & & $-0.99\%$     \\
             &   &        &           &          &            &                        &   $H_P: 3.937 \times 10^{-9}$  & $0.91$ &    $-1.68\%$    \\
             &   &        &           &          &            &                        &   $B_T: 3.724 \times 10^{-9}$  & $3.96$ &    $-7.00\%$    \\
             &   &        &           &          &            &                        &   $V_T: 3.976 \times 10^{-9}$  & $2.99$ &    $-0.72\%$    \\
             &   &        &           &          &            &                        &   $J:   3.896 \times 10^{-9}$  & $4.37$ &    $-2.71\%$    \\
             &   &        &           &          &            &                        &   $H:   4.175 \times 10^{-9}$  & $5.42$ &    $4.25\%$    \\
             &   &        &           &          &            &                        &   $K_S: 3.983 \times 10^{-9}$  & $5.04$ &    $-0.53\%$    \\
             &   &        &           &          &            &                        &   $B:   3.921 \times 10^{-9}$  & $2.89$ &    $-2.08\%$    \\
             &   &        &           &          &            &                        &   $V:   4.002 \times 10^{-9}$  & $1.56$ &    $-0.07\%$    \\
             &   &        &           &          &            &                        &   $R_C: 3.994 \times 10^{-9}$  & $1.51$ &    $-0.26\%$    \\
             &   &        &           &          &            &                        &   $I_C: 4.042 \times 10^{-9}$  & $2.06$ &    $0.94\%$    \\
\hline
\end{tabular}
\end{table*}
\setcounter{table}{1}
\begin{table*}
\centering
\caption{Continued}
\begin{tabular}{cccccccrrr}
\hline 
 Star    & Tier & $\teff$~(K) &  $\logg$  &  $\feh$  &  $E(B-V)$  & $\fbol$ \tiny{CALSPEC} &  $\fbol$ \tiny{MARCS} & $\sigma(\%)$ &  $\Delta(\%)$ \\
\hline
HD160617     & 3 & $6048$   & $3.73$    &  $-1.78$ &  $0.005$   &  $ 9.510 \times 10^{-9}$  &  $<> :  9.406 \times 10^{-9}$  & & $-1.09\%$     \\
             &   &        &           &          &            &                        &   $H_P: 9.482 \times 10^{-9}$  & $1.14$ &  $-0.30\%$    \\
             &   &        &           &          &            &                        &   $B_T: 9.520 \times 10^{-9}$  & $3.62$ &  $0.11\%$    \\
             &   &        &           &          &            &                        &   $V_T: 9.286 \times 10^{-9}$  & $2.00$ &  $-2.35\%$    \\
             &   &        &           &          &            &                        &   $J:   9.582 \times 10^{-9}$  & $3.62$ &  $0.75\%$    \\
             &   &        &           &          &            &                        &   $H:   9.390 \times 10^{-9}$  & $5.65$ &  $-1.27\%$    \\
             &   &        &           &          &            &                        &   $K_S: 9.261 \times 10^{-9}$  & $5.27$ &  $-2.62\%$    \\
             &   &        &           &          &            &                        &   $B:   9.390 \times 10^{-9}$  & $3.29$ &  $-1.27\%$    \\
             &   &        &           &          &            &                        &   $V:   9.390 \times 10^{-9}$  & $1.66$ &  $-1.27\%$    \\
             &   &        &           &          &            &                        &   $R_C: 9.372 \times 10^{-9}$  & $1.50$ &  $-1.45\%$    \\
             &   &        &           &          &            &                        &   $I_C: 9.390 \times 10^{-9}$  & $2.02$ &  $-1.27\%$    \\
\hline
\end{tabular}
\begin{minipage}{1\textwidth}
$\fbol$ {\tiny MARCS} is the bolometric flux recovered in different bands:
$H_P$ Hipparcos, $B_TV_T$ Tycho, $JHK_S$ 2MASS and $BV(RI)_C$ Johnson-Cousins.
For the latter system we have used the {\tt ubvri12} transformations (see 
Paper I). $<>$ is the average of all systems listed
for a given star. $\sigma(\%)$ is the percent uncertainty in the bolometric flux
recovered from a single band, and estimated from MonteCarlo as described in
text. $\Delta (\%)$ is the percent difference between the CALSPEC bolometric
flux and that recovered from our bolometric corrections. 
In case of reddening, also the CALSPEC flux has been corrected by
the same amount, with the same extinction law adopted in our tables of
bolometric corrections.
\end{minipage}
\end{table*}

\subsubsection{Tier 3}

Finally, we list stars which are affected by reddening as Tier 3. These stars
are further away, generally at $\sim 100$~pc, and they all happen to be
metal-poor. In this case,
we apply the same reddening corrections to the CALSPEC spectra, as well as to the
observed photometry. As for stars in the Tier 2 category, all of these stars have
STIS observations between $1700$ and $10000$~\AA, with extensions
longward of $10000$~\AA\ derived from model atmospheres. We adopt stellar parameters
from \cite{mcr10}, who determined the reddenings of these stars
from interstellar Na\,\RN{1}\,D lines. Also, the $\teff$ scale for stars in
both Tier 2 \citep{c11} and Tier 3 \citep{mcr10} is based on the same
implementation of the infrared flux method \citep{c10}. 

We use synthetic photometry of an archetypal metal-poor star ($\teff=6000$K,
$\logg=4.0$, $\feh=-2.0$) to assess the impact of reddening on bolometric
flux. A variation of $0.01$ in $E(B-V)$ affects the recovered bolometric flux
by an amount which varies between 0.3--0.8 percent in the infrared, to around
2--3 percent in the optical. On average we recover the CALSPEC bolometric
fluxes to within $\simeq 1$ percent, but single bands can occasionally return
differences as large as
5--7 percent. Also in this case, differences with respect to CALSPEC
bolometric fluxes are usually within the MonteCarlo uncertainties.

\subsubsection{Regarding the precision and accuracy of bolometric fluxes}

In the above comparisons, we have estimated the mean uncertainty in recovering
bolometric fluxes using MonteCarlo studies that take into account the uncertainties
in the input parameters and observed photometry. The mean uncertainty in the
bolometric fluxes obtained from our tables increases from around 2 percent for
the star in Tier 1, to 3 percent for stars in Tier 3, reflecting the lower
quality of the input parameters. However, comparisons with CALSPEC fluxes
suggest that the precision is usually better than those derived from the
MonteCarlo analyses.

For a given band, the average difference between the fluxes we recover and the
CALSPEC observations is $\pm1.1$ percent at Tier 1, increasing to
$\pm1.7$\ percent and $\pm1.8$\ percent for stars in the Tier 2 and Tier 3
categories, respectively. When combining the results for all available
bands, the mean difference varies from $-0.2$ percent for the Tier 1 star,
to $\sim -0.7$ percent for stars in the Tier 2 and 3 categories. This difference
is clearly systematic, in the sense that our fluxes are usually smaller.
These findings could be modified by changing the adopted value of
$L_\odot$, but it should also be kept in mind that the systematic uncertainty
in CALSPEC spectrophotometry is at the 1 percent level \citep{bohlin14}. Hence,
we conclude that, when combining the results from different bands, it is
feasible to recover the bolometric flux to within 1 percent, below which we
are limited by the fundamental uncertainty of the current absolute flux scale. 
\begin{figure*}
\begin{center}
\includegraphics[width=0.9\textwidth]{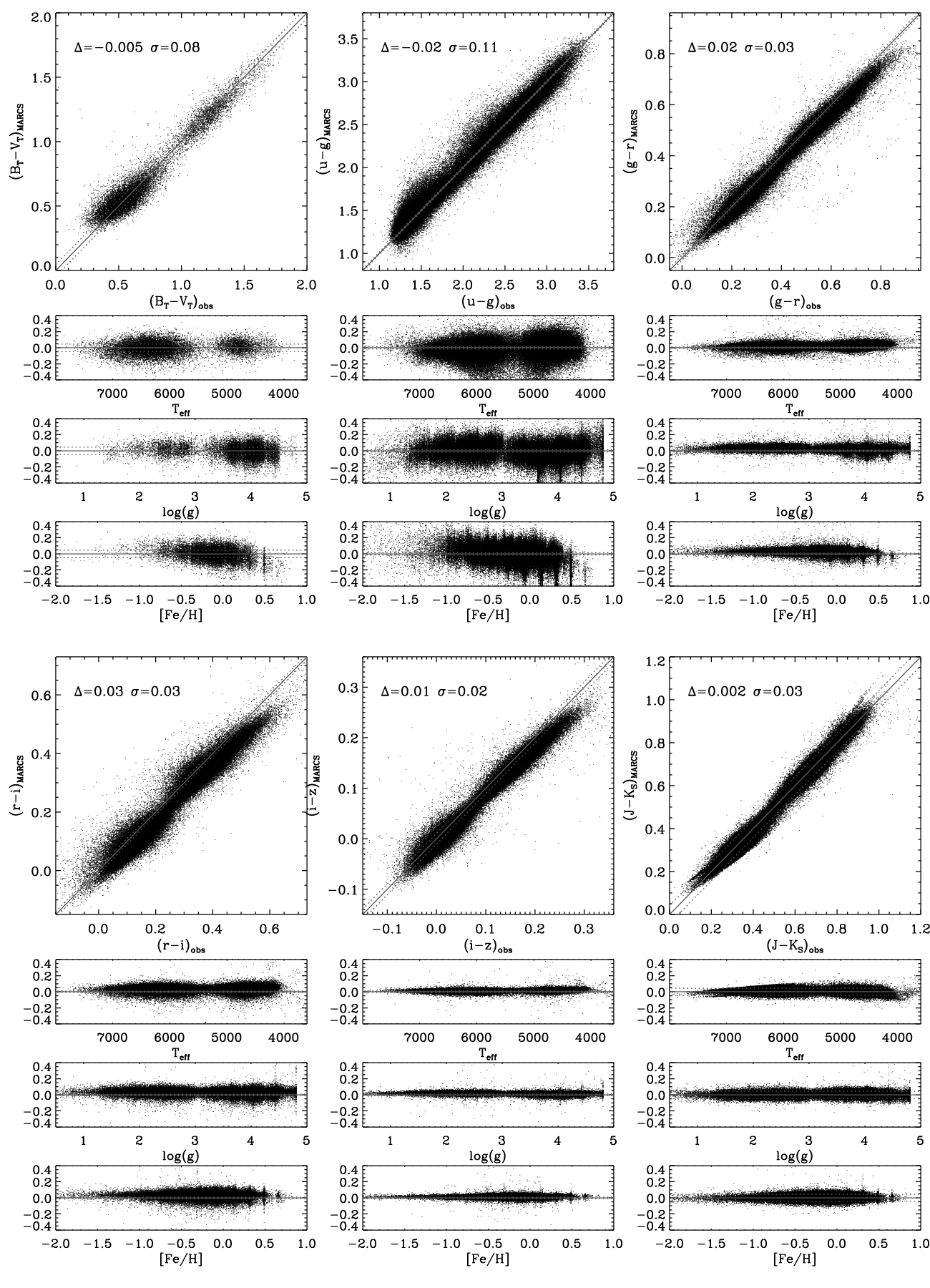}
\caption{Comparison between observed and synthetic MARCS colours interpolated
  from our tables at the IRFM $\teff$, $\logg$, $\feh$ and rescaled $E(B-V)$
  of each RAVE DR5 star. Continuous gray line is the one-to-one relation,
  whereas dotted
  lines indicate the mean scatter arising solely from observed photometric
  errors. $\Delta$ and $\sigma$ are the mean colour index difference (observed
  minus synthetic) and scatter in mag.}\label{f:cc}
\end{center}
\end{figure*}

\subsection{Testing colour indices}

Another way to test the performance of our synthetic photometry is by comparing
various combinations of synthetic and observed colours for stars of known
stellar parameters. In order to do this, we cross-matched RAVE DR5 \citep{kunder}
with the Tycho, 2MASS and SkyMapper catalogues. RAVE DR5 provides stellar
parameters for nearly $0.5$ million stars in the southern hemisphere: we retained
only those stars with the best determined stellar parameters ({\tt ALGO\_CONV$=0$,
  c1=c2=c3=n}). We rescaled the reddening from \cite{sfd98} as described in
\citet[][which in fact provides overall good agreement between photometric and
spectroscopic $\teff$]{kunder}. In addition, to avoid regions with severe
extinction, we considered only stars at Galactic latitudes $|b|>10^\circ$ and
excluded stars with rescaled $E(B-V) \geqslant 0.1$. Further cuts on
photometric quality were also applied, depending on the filters
(see captions to figures), typically yielding $>100,000$\ usable stars.
We compared observed and synthetic colours by plotting stars in
different ranges of $\logg$. Stars with lower surface gravities have higher
intrinsic luminosities, and thus in RAVE (grossly magnitude limited) they
span larger distances. The metallicity gradients in the Galactic disc
\citep[e.g.,][]{bsp13,bsp14} explain why the bulk metallicity
decreases with decreasing $\logg$.

MARCS synthetic colours perform overall well across a wide range of filters
and stellar parameters, as showcased in Figure \ref{f:col4} through
\ref{f:col3}. They are
able to predict the main observed features, although they fail in the $u-g$
vs $g-r$ plane at the lowest gravities and hottest $\teff$. To better quantify
these findings, we generated synthetic colours for each RAVE DR5 star that
passed our quality and photometric cuts. This is shown in Figure
\ref{f:cc}, which compares observed and synthetic colours as function of
stellar parameters. Again, the overall agreement between observed and synthetic
photometry is quite remarkable, with mean offsets that are always within the
scatter and never exceeding a few hundredths of a mag. In particular, the Tycho and
2MASS colours have nearly no offsets, indicating that the synthetic colours in these
two systems are extremely well standardised (both observationally and
theoretically). The overall agreement for the SkyMapper colours is equally
remarkable, especially considering that, at the present time,
the exact standardisation of observed SkyMapper photometry to
the {\tt AB} system is still work in progress. 

It can immediately be seen that indices involving bluer colours ($B_T-V_T$
and $u-g$), while well centred at the solar $\feh$, are increasingly offset at
higher and lower metallicities. This is likely due to the fact
that, at short wavelengths, synthetic spectra depend critically on the adopted
opacities and microturbulence (see the discussion in Paper I). The SkyMapper $u$
filter is also sensitive to gravity, and it is apparent the predicted colours
depart from the observed ones for $\logg < 1.5$. There is also a tendency for
the synthetic colours to be bluer than observed at $\teff \lesssim 4500$~K,
which could indicate missing molecular opacities. 

\subsection{Star Cluster Colour-Magnitude Diagrams}\label{subsec:cmds}

The colour-magnitude diagrams (CMDs) of open and globular star clusters provide
especially good constraints on properties of stellar models for lower mass
stars, including their colours, because the member stars have very close to the
same age and initial chemical abundances \citep[modulo the presence of
  multiple populations, see e.g.,][for a review]{gcb}.
The CMDs therefore describe how the
colours (and effective temperatures) of stars vary with gravity at a fixed age
and metallicity.  Even though current stellar models appear to be able to
reproduce the principal photometric sequences of star clusters quite well (see,
e.g., \citealt{vbl13}, \citealt{vbf14}; Paper I), the predicted $\teff$\ scale
is subject to uncertainties associated with, e.g., convection theory, the
atmospheric boundary condition, and the treatment of diffusive processes, while
synthetic colours depend sensitively on $\teff$, the temperature structures of
model atmospheres, the treatment of blanketing, etc.  Consequently, one should
not expect to find perfect agreement between isochrones that have been
transformed to various observational planes and observed CMDs.

Still, the consistency of predicted colour--$\teff$\ relations can be tested by
determining if the same interpretation of the data (including discrepancies
between theory and observations) is found on many different CMDs.  For instance,
in the study of $UBV(RI)_C$, Sloan ($ugriz$), and {\it HST}-ACS photometry of
the globular cluster (GC) M\,5 that was presented in Paper I (see their 
figs.~9--11), the red giant branch (RGB) of the isochrone that was fitted to the
observations was consistently located along the red edge of the observed
distribution of giants in all of the different CMDs that were considered.  The
most likely explanation of this offset is that the predicted temperatures along
the RGB are somewhat too cool.  Moreover, with the exception of the
[$(F606W-F814W),\,F606W$]- and [$(g-r),\,r$]-diagrams, the same isochrone
provided an excellent match to the turnoff (TO) and upper main sequence (MS)
observations.  Here again, very encouraging consistency was found, though it
would appear that either the {\it HST} and Sloan observations (in the $F606W$
and/or the $F814W$ passbands and in the $g$ and/or $r$ filters, respectively),
or the corresponding bolometric correction transformations presented in
Paper I, suffer from small zero-point errors.

\begin{figure}
\begin{center}
\includegraphics[width=0.48\textwidth]{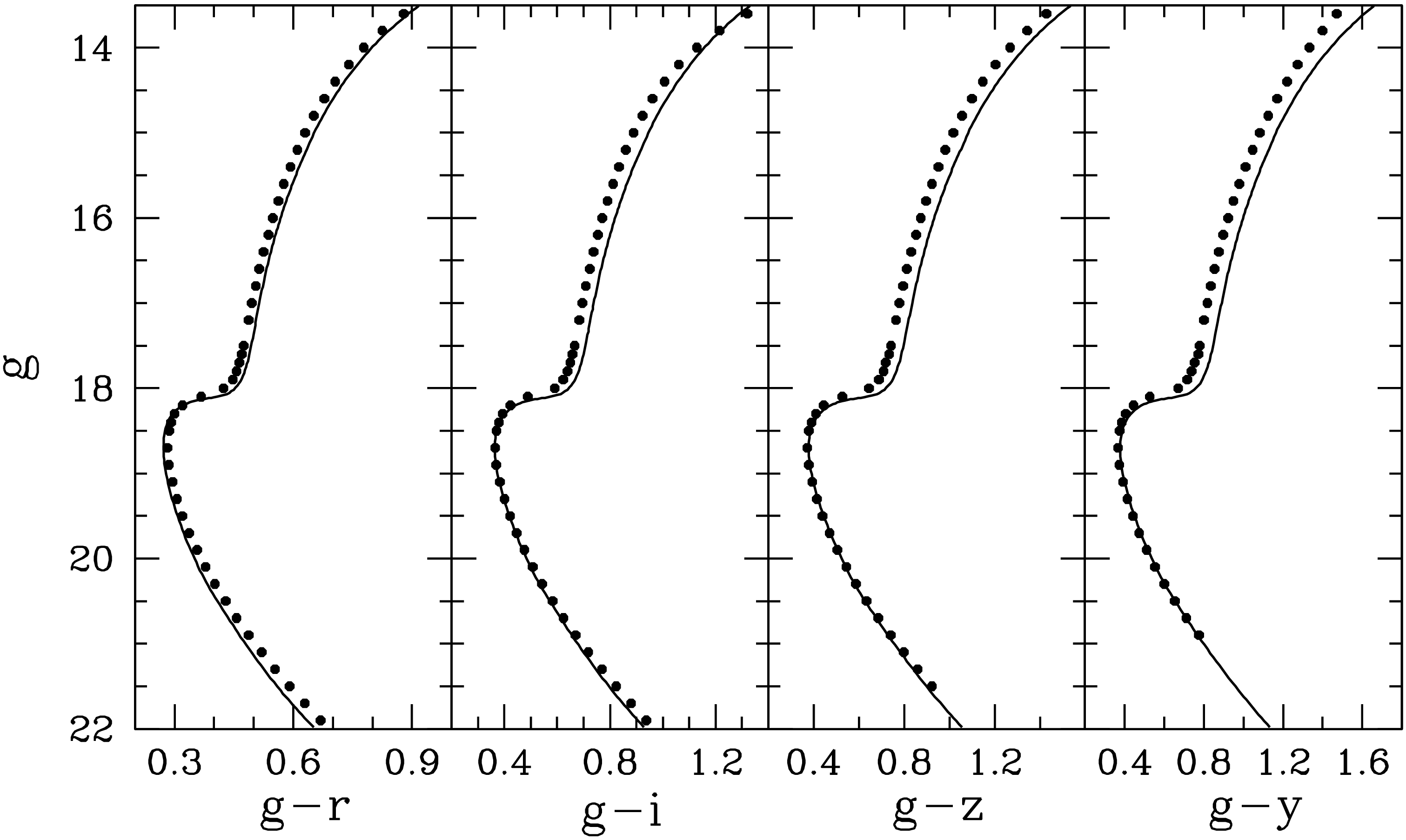}
\caption{Overlay of an $11.75$ Gyr isochrone for [Fe/H] $= -1.33$, [$\alpha$/Fe]
$= 0.4$, and $Y=0.25$ (solid curve) to the principal photometric sequences of
M$\,$5 (small filled circles) that were derived by \citet{bfs14} from their
Pan-STARRS1 $grizy$ observations.  The transformation of the isochrone to
the observed planes assumes $E(B-V) = 0.032$ (e.g., \citealt{sf11}) and
$(m-M)_0 = 14.35$.}\label{fig:m5}
\end{center}
\end{figure}

This investigation affords us with the opportunity to extend the aforementioned
analysis of star cluster CMDs to the Pan-STARRS1 photometric system given
that \citet{bfs14} have published well-defined fiducial sequences for
the same GC (M$\,$5) and open clusters (M$\,$67 and NGC$\,$6791) that were
considered in Paper I.  If, as adopted previously, an isochrone for $11.75$ Gyr,
[Fe/H] $= -1.33$, [$\alpha$/Fe] $= 0.4$, and $Y=0.25$ is overlaid onto the
Pan-STARRS1 fiducials assuming the same cluster parameters ($E(B-V) = 0.032$
and a true distance modulus $(m-M)_0 =14.35$), we obtain the results shown in
Figure \ref{fig:m5}.  As before, the observed RGB is somewhat bluer than the
predicted giant branch, though both loci have nearly the same slope, while the
upper MS observations are generally quite well matched by the models ---
except in the case of the [$(g-r),\,g$]-diagram, where the isochrone is too
blue by $\sim 0.02$ mag. Since no such problem is apparent in the other CMDs
that involve the $g$ passband, it would appear that there is a small
zero-point problem with either the $r$ photometry or the $BC_r$ predictions.
Nevertheless, for the most part, the selected isochrone provides nearly
identical fits to Pan-STARRS1 observations as to $BV(RI)_C$ and Sloan data
(see Paper I). This indicates that the transformations to the different
photometric systems are highly consistent with one another.

\begin{figure}
\begin{center}
\includegraphics[width=0.48\textwidth]{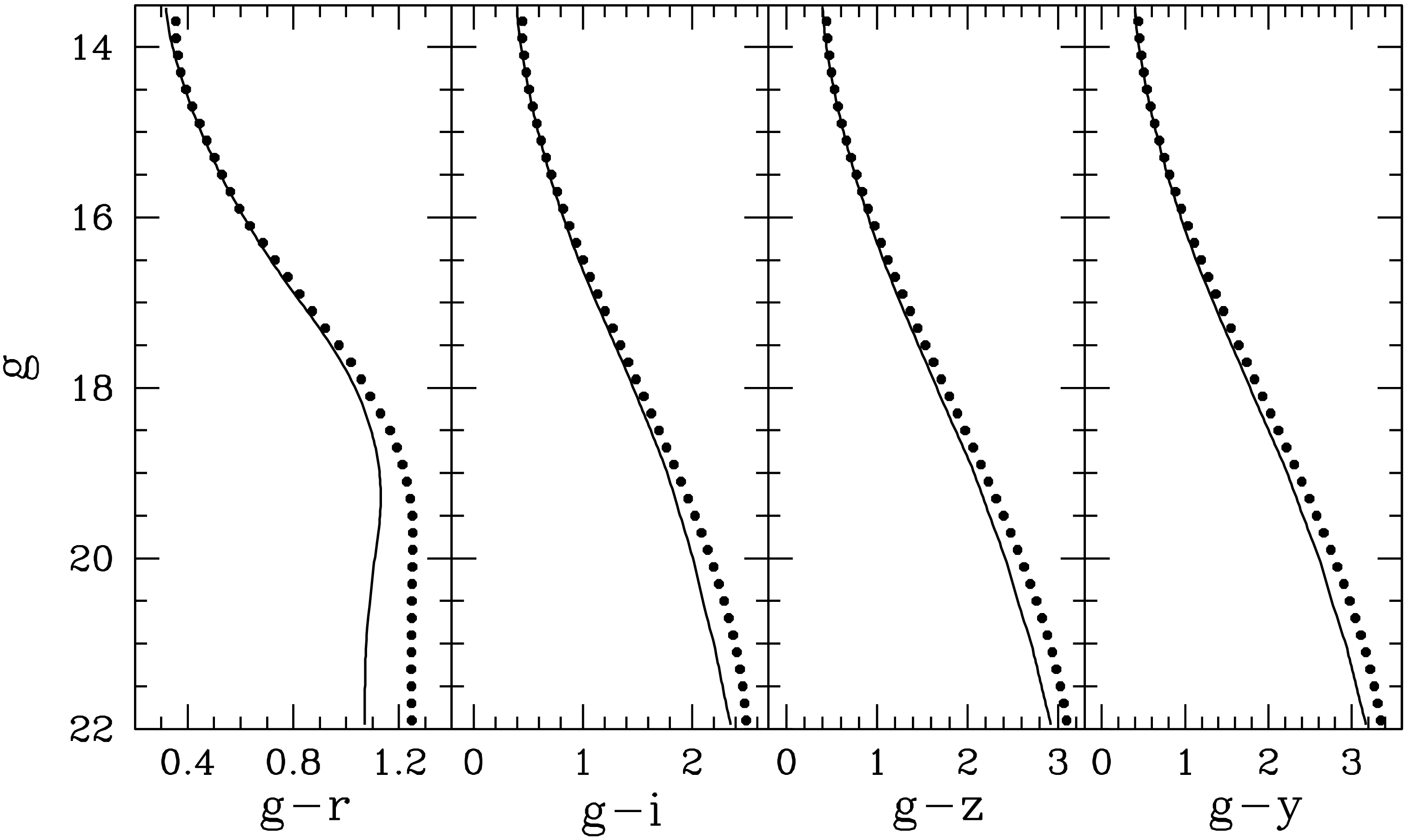}
\caption{Similar to the previous figure; in this case, a $4.3$ Gyr isochrone for
[Fe/H] $ = 0.0$ and $Y=0.255$ has been overlaid on the main-sequence fiducial
of M\,67 derived by \citet{bfs14}, assuming $E(B-V) = 0.03$ (e.g.,
\citealt{sf11}) and $(m-M)_0 = 9.60$.}\label{fig:m67}
\end{center}
\end{figure}

The fact that bluer colours are more difficult to match than those derived from
filters at longer wavelengths is almost certainly due primarily to deficiencies
in the modelling of stellar atmospheres and synthetic spectra, as is the
increased difficulty of explaining the colours of cooler stars.  For instance,
figure 14 in Paper I shows that a 4.3 Gyr, solar abundance isochrone provides a
very good fit to the MS of the open cluster M$\,$67 down to $V \sim 16.7$ on the
[$(B-V),\,V$]-diagram, as compared with $V \sim 19$ on the
[$(V-K_S),\,V$]-diagram.  Moreover, at fainter magnitudes, the differences
between the predicted and observed colours typically increase with decreasing
$\teff$.  The same behaviour is found (see Figure \ref{fig:m67}) when the same
isochrone is fitted to
Pan-STARRS1 photometry of M$\,$67 (\citealt{bfs14}) on the assumption of
$E(B-V) = 0.03$\ and $(m-M)_0 = 9.60$\ (as adopted in Paper I).  On the one
hand, it is very gratifying to find that this isochrone provides a superb fit
to the upper MS, irrespective of the filter system that was employed.
On the other hand, it is clear that the predicted colours of lower-MS stars are
too blue by $\gta 0.2$\ mag at $g > 20$.  (We note that such CMD comparisons
as those plotted in Figure \ref{fig:m67} could be used to derive empirical
corrections to the purely synthetic colours.)

\begin{figure}
\begin{center}
\includegraphics[width=0.48\textwidth]{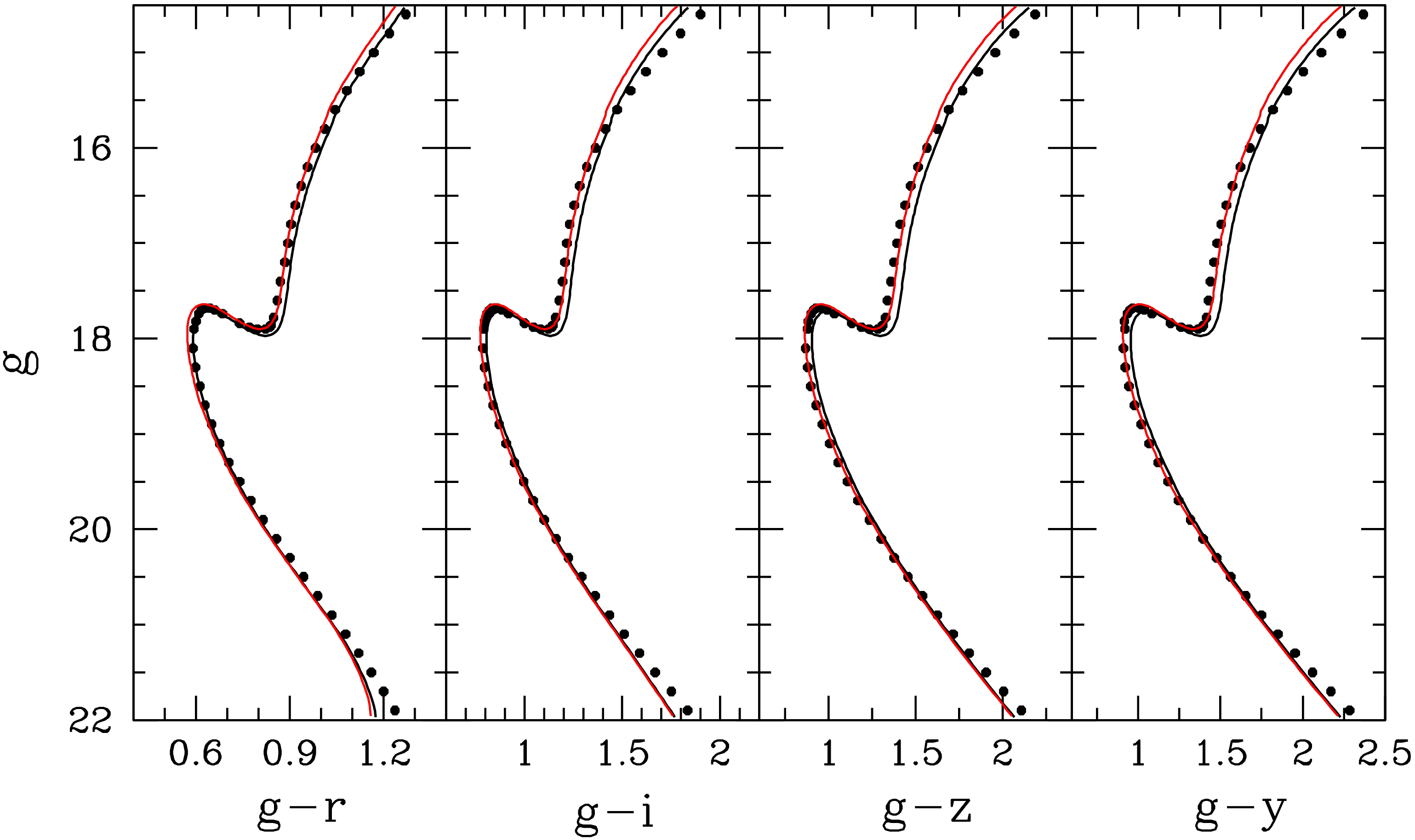}
\caption{Similar to the previous two figures; in this case, an isochrone for
$8.5$ Gyr, [Fe/H] $= +0.30$, and $Y=0.28$ has been overlaid on the main-sequence
fiducial of NGC$\,$6791 derived by \citet{bfs14}, assuming $(m-M)_0 = 13.05$ and
$E(B-V) = 0.16$ (black curve) or $0.14$ (red curve).}\label{fig:6791}
\end{center}
\end{figure}

Although an isochrone for 8.5 Gyr, [Fe/H] $= +0.30$, [$\alpha$/Fe] $= 0.0$, and
$Y=0.28$ provides a very good fit, in an absolute sense, to many of the CMDs
of NGC$\,$6791 that can be constructed from $BVI_CJ$ and Sloan $ugriz$
photometry (see figures 12 and 13 in Paper I), it does not reproduce the
fiducial sequences that were derived from Pan-STARRS1 observations by
\citet{bfs14} quite as well.  As shown in Figure \ref{fig:6791}, the
predicted TO and RGB tend to be somewhat too red if $E(B-V) = 0.16$ and
$(m-M)_0 = 13.05$ (as adopted in Paper I, which discusses recent
determinations of the cluster parameters), whereas no such discrepancies were
found previously.  Much better consistency can be obtained if $E(B-V) = 0.14$
(the red curve) or a slightly larger value, likely indicating a small
zero-point offset ($0.01-0.02$ mag) between Pan-STARRS1 photometry and
Johnson-Cousins-Sloan observations of NGC$\,$6791.  In fact, a small
difference in the same sense was found by \citet[see their figure 5]{bfs14}
when they compared their
fiducial sequences with those obtained by transforming the Sloan CMDs given by
\citet{ajc08} to the Pan-STARRS1 system. Although further work is needed to
fully understand this discrepancy, this level of consistency is still quite
satisfactory given that observations typically involve zero-point
uncertainties close to $0.01$ mag \citep[see e.g.,][]{st05}.

In fact, it is remarkable that the Pan-STARRS1 transformations yield
colours that provide fully consistent interpretations of the data (to within
$0.01$ mag) as the transformations considered in Paper I over a wide range in
metallicity and gravity --- especially for stars within $\sim \pm 2$ mag of
the turnoff.

\section{Conclusions}

In this paper we have continued our effort to provide reliable and well tested
synthetic colours and bolometric corrections from the MARCS library for various
combinations of $\teff, \logg, \feh$ and $\aFe$. We have extended our previous
investigation (Paper I) to include many of the systems underpinning large
photometric surveys in the optical (Hipparcos/Tycho, Pan-STARRS1 and SkyMapper),
and we plan to add {\it Gaia} photometry once the transmission curves and
standardisation details become available. Together with 2MASS and SDSS
(included in Paper I, among other systems), these surveys are fostering a wide
variety of studies in stellar and Galactic astronomy. However to reap the
maximum returns, properly standardised, reliable and well tested stellar
synthetic colours are needed to interpret the observations.

Encouragingly, our study of star clusters has demonstrated that isochrones
generally provide very similar fits to observed CMDs in different photometric
systems under the same assumptions concerning the age, reddening and distance
modulus. Small adjustments at the level of $0.01-0.02$ mag are sometimes
needed to obtain fully consistent interpretations of the data, but this appears
to be an exception more than the rule. This indicates that overall we have
achieved our goal of providing synthetic colours that are homogeneously
standardised across different system to within $\sim 0.01$ mag, which is also
the zero-point limit of typical ground-based observations. 

Once we trust our zero-points, we can have meaningful discussions concerning the
performance of the MARCS synthetic photometry. We have done so by computing
synthetic colours for more than a hundred thousand stars from RAVE DR5, and
comparing them with the observed colours.
Overall, we find very satisfactory agreement of optical and infrared
colours for most of the parameter space that has been explored. However,
at short wavelengths, models have some difficulties predicting colours for
stars with gravities roughly below $\logg=1.5$, and while good agreement between
observed and synthetic colours is always found close to the solar metallicity,
differences become evident in the direction of both lower and higher
$\feh$. Irrespective of the filters that are used, discrepancies always begin to
appear as the temperature drops toward the coolest $\teff$\ values that we have
considered. We recall from Paper I that for the blue and ultraviolet spectral
region (in particular for the coolest and most metal rich stars) synthetic
colours show dependence on microturbulence, which is fixed to $2\,\kms$ in this
investigation. (Note that this investigation has focused on FGK type stars.
Although we predict the BCs for M dwarfs and giants, their performance is not
as thoroughly tested as for earlier spectral types.)

{\it HST} spectrophotometry has enabled us to perform the most stringent test
yet on the level at which bolometric fluxes can be recovered from our tables of
synthetic BCs. For FG dwarfs of known input stellar parameters, we
conclude that, on average, bolometric fluxes can be recovered to within about
2 percent from our computed bolometric corrections in a single band, and that
this uncertainty is usually halved when combining the results from more bands.  
To facilitate further investigations of MARCS colours, we have provided tables
of BCs and suitable computer programs to interpolate in them for any
combination of $\teff, \logg, \feh$ and $\aFe$\ that is contained within the
MARCS library.  The most obvious application of these data and interpolation
codes, which are described in the Appendix, is to transpose isochrones to many
different CMDs.  Among the photometric systems that have been considered (here
and in Paper I), we have included synthetic photometry for the imaging cameras
on {\it JWST}. While the standardisation of real data from {\it JWST} may turn
out to be somewhat different from what we have assumed, our programs make it
possible to obtain a first estimate of the appearance of stellar populations
through the eye of {\it JWST} and, therefore, to assist with the planning of
future observations. 

\section*{Acknowledgments}
LC gratefully acknowledges support from the Australian Research Council (grants
DP150100250, FT160100402), while the contributions of D.A.V~to this project
were supported by a Discovery Grant from the Natural Sciences and Engineering
Research Council of Canada. We thank Andrea Kunder for useful advice on the
use of RAVE flags, Edouard Bernard for providing machine-readable versions of
the globular cluster fiducials in Pan-STARRS1, George Rieke for helpful
correspondence on JWST filters, and Mike Bessell for comments.
The national facility capability for SkyMapper has been funded through ARC LIEF
grant LE130100104 from the Australian Research Council, awarded to the
University of Sydney, the Australian National University, Swinburne University
of Technology, the University of Queensland, the University of Western
Australia, the University of Melbourne, Curtin University of Technology, Monash
University and the Australian Astronomical Observatory. SkyMapper is owned and
operated by The Australian National University's Research School of Astronomy
and Astrophysics. The survey data were processed and provided by the SkyMapper
Team at ANU. The SkyMapper node of the All-Sky Virtual Observatory (ASVO) is
hosted at the National Computational Infrastructure (NCI). Development and
support the SkyMapper node of the ASVO has been funded in part by Astronomy
Australia Limited (AAL) and the Australian Government through the Commonwealth's
Education Investment Fund (EIF) and National Collaborative Research
Infrastructure Strategy (NCRIS), particularly the National eResearch
Collaboration Tools and Resources (NeCTAR) and the Australian National Data
Service Projects (ANDS). This publication makes use of data products from the
Two Micron All Sky Survey, which is a joint project of the University of
Massachusetts and the Infrared Processing and Analysis Center/California
Institute of Technology, funded by the National Aeronautics and Space
Administration and the National Science Foundation. Funding for RAVE
(www.rave-survey.org) has been provided by institutions of the RAVE
participants and by their national funding agencies.

\bibliographystyle{mn2e}
\bibliography{refs}

\newpage

\appendix

\section[]{Interpolation routines}\label{appA}

The Appendix in Paper I provides a detailed and quite thorough description of
the library of bolometric corrections (BCs) that has been developed for several
of the most widely used broad-band photometric systems (Johnson-Cousins, SDSS,
2MASS, and {\it HST}-ACS, {\it HST}-WFC3, calibrated to {\tt AB}, {\tt ST},
and {\tt VEGA}
zero-points).  It contains data for more than 40 filters, derived from MARCS
model atmospheres \citep[][same as used here]{g08} that encompass wide ranges
in gravity
($-0.5 \le \logg \le 5.5$), effective temperature ($2600 \le \teff
\le 8000$~K), iron abundance ($-4.0 \le$ [Fe/H] $\le +1.0$), and
$\alpha$-element abundances ($-0.4 \le$ [$\alpha$/Fe] $\le +0.4$) --- though
the coverage is not complete (e.g., BCs are not available for the full range
in $\teff$ at all gravities; see figure A2 in Paper I) and the range in [Fe/H]
varies somewhat for different values of [$\alpha$/Fe] (see below).  Moreover,
BCs have been generated for $E(B-V) = 0.0$, 0.12, 0.24, $\ldots$, 0.72 so
that, by interpolation in these results, one can take the effects of
interstellar reddening into account. This library is contained in the file
{\tt BCtables.tar.gz}, whereas {\tt BCcodes.tar} contains FORTRAN
interpolation routines. 

In this investigation, we have created a second library ({\tt BCtables2.tar.gz})
comprised of analogous transformations for the SkyMapper, Hipparcos/Tycho,
Pan-STARRS1, and {\it JWST}-NIRCam/MIRI systems. {\tt BCtables2.tar.gz},
together with {\tt BCtables.tar.gz} from Paper I can be use with the new and
updated set of interpolation routines described below ({\tt BCcodes.tar}),
which supersede those provided in Paper I\footnote{All material is freely
  available online through CDS, and kept up-to-date on GitHub \href{https://github.com/casaluca/bolometric-corrections}{github.com/casaluca/bolometric-corrections}}.
Because most users will be interested in only a small subset of the predicted
BCs, the supplied FORTRAN codes will generate tables of reddening-corrected
BCs, assuming just one user-selected value of $E(B-V)$, for up to a maximum of
5 different filters.  The filter selection is controlled by
{\tt selectbc.data}, which has already been described in considerable detail
in Paper I.  In fact, the only substantive change that has been made to this
file is an extension of the menu options to include the additional photometric
systems and associated filters that are the subject of this paper.  

\begin{figure}
\begin{center}
\includegraphics[width=0.45\textwidth]{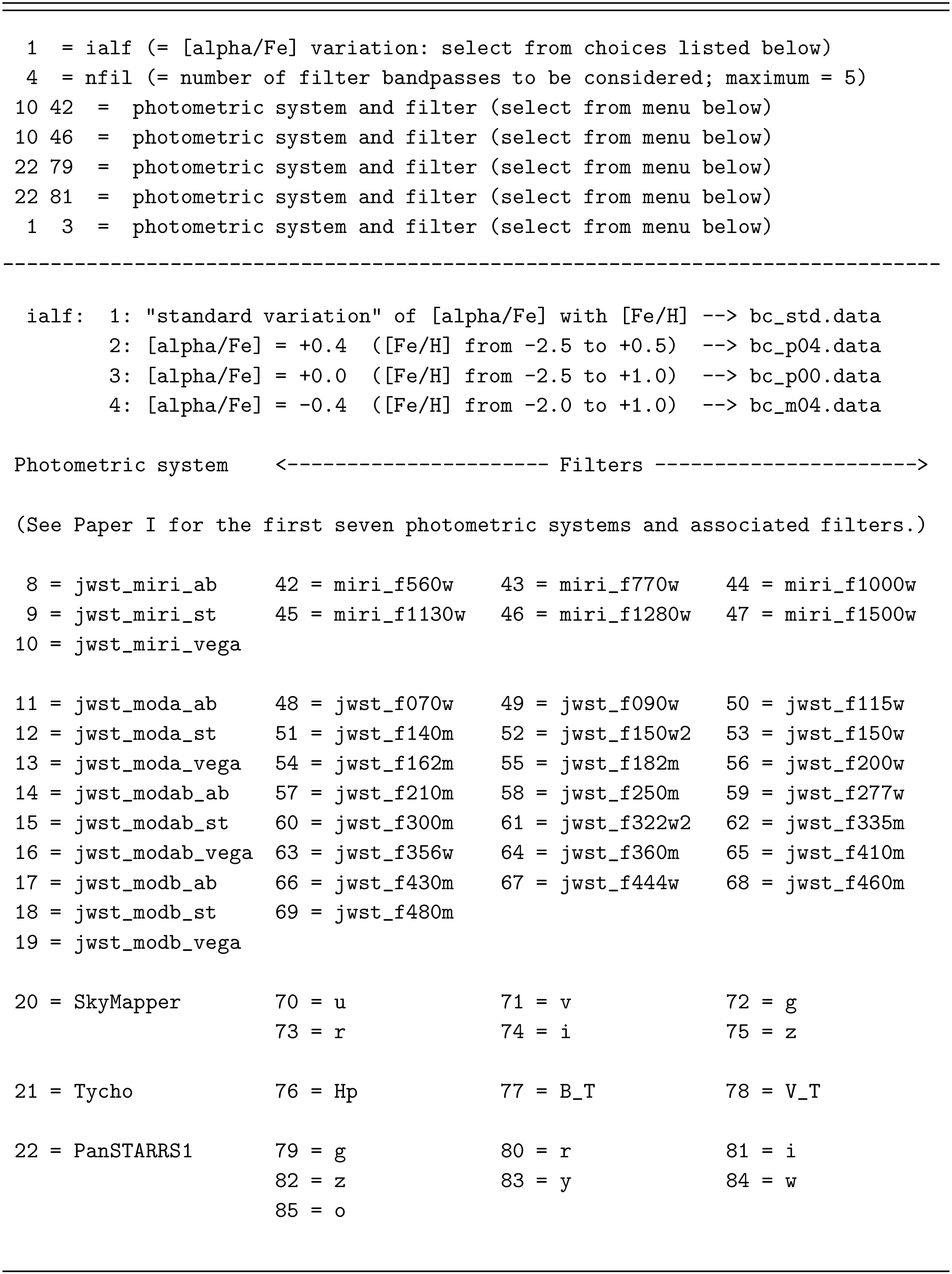}
\caption{Listing of the parts of {\tt selectbc.data} relevant to the photometric
systems and associated filters considered in this study.  (The menu options
relevant to the photometric systems considered in Paper I have been omitted, but
see figure A3 in that study.)  Depending on the values of the integer parameters
in the seven lines just above the dashed line, the supplied FORTRAN codes will
produce BC tables for up to 5 filters of interest and one of four possible
variations of [$\alpha$/Fe] with [Fe/H] (see the text for a detailed
explanation.)}
\label{fig:figA1}
\end{center}
\end{figure}

As shown in Figure \ref{fig:figA1}, there are now 22 photometric systems (or
variations
thereof) and a total of 85 different filters in the combined library.  (Because
the part of the menu concerning the first seven photometric systems already been
discussed in Paper I, it has not been reproduced here.)  According to the text
below the dashed line, we see, for instance, that the integers 8, 9, and 10 in
the left-hand column identify synthetic {\it JWST}-MIRI photometry,
calibrated, in
turn, to the {\tt AB}, {\tt ST}, and {\tt VEGA} systems.  The only filters that
are relevant to these cases are F560W to F1500W, which have integer
identifications ranging from 42 to 47, respectively.  Similarly, BCs have been
computed for 22 {\it JWST}-NIRCam filters (F070W, F090W, $\ldots$, F480M) using
module A ({\tt moda}), module B ({\tt modb}), and the average of the two, AB
({\tt modab}. See discussion in Section \ref{sec:nircam}). Since each of these 3
cases has also been calibrated to the {\tt AB}, {\tt ST}, and {\tt VEGA}
systems, NIRCam data has been subdivided into the 9 possibilities that
are identified by the integers 11 to 19 in the left-hand column. 

Thus, to obtain any BC tables of interest, it is simply a matter of specifying
the number of filters ({\tt nfil}) on the second line of {\tt selectbc.data},
followed by pairs of integers on successive lines to identify those filters.
In the example given in Figure \ref{fig:figA1}, {\tt nfil} $= 4$ and the
selected filters are MIRI F560W, F1280W (calibrated to {\tt VEGA} zero-points)
and Pan-STARRS1 $g, i$. The first line of {\tt selectbc.data} contains the
parameter {\tt ialf}, which determines whether the BCs are to assume the
``standard variation" of [$\alpha$/Fe] with [Fe/H] ({\tt ialf} $= 1$), or
[$\alpha$/Fe] $= +0.4$, 0.0, or $-0.4$ for the ranges in [Fe/H] that are given
in the menu ({\tt ialf} $= 2$, 3, or 4, respectively).  The standard variation
assumes that [$\alpha$/Fe] $= +0.4$ for $-4.0 \le$ [Fe/H] $< -1.0$,
[$\alpha$/Fe] $= -0.4\,\times$[Fe/H] for $-1.0 \le$ [Fe/H] $\le 0.0$ (i.e., a
linear
decrease in [$\alpha$/Fe] from $+0.4$ at [Fe/H] $= -1.0$ to 0.0 at [Fe/H]
$= 0.0$), and [$\alpha$/Fe] $=0.0$ for $0.0 \le$ [Fe/H] $\le +1.0$. 

\begin{figure}
\begin{center}
\includegraphics[width=0.45\textwidth]{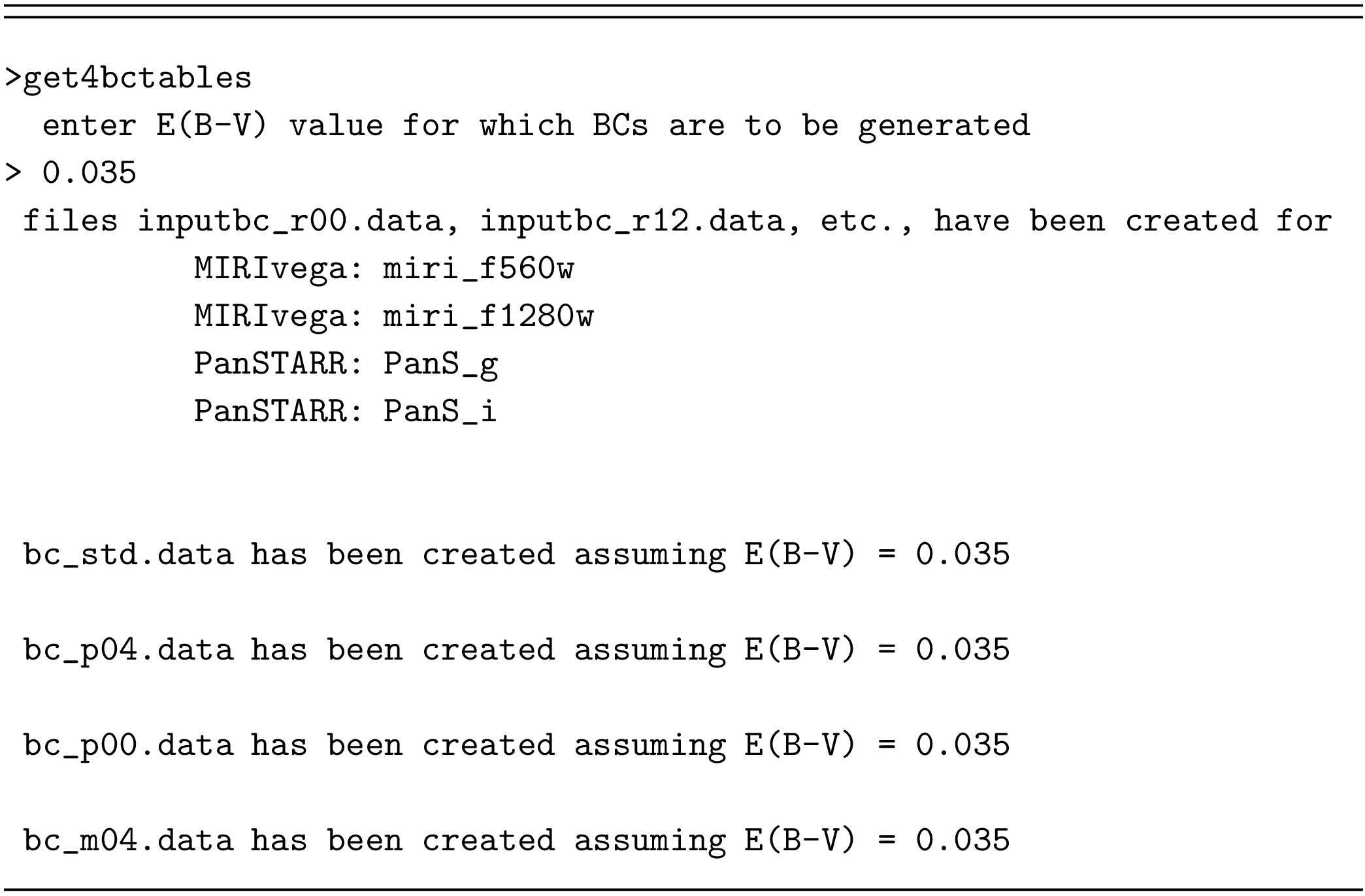}
\caption{Sample execution of {\tt get4bctables}.  When the name of this 
executable module is entered at the prompt ({\tt >}), the code notifies both
the filters selected by the user and the four data files that have been
generated containing the reddening-corrected BC tables for those filters.}
\label{fig:figA2}
\end{center}
\end{figure}

The procedure to compute BCs for different variations of [$\alpha$/Fe] was
somewhat cumbersome in Paper I. We now provide a new program named
{\tt get4bctables.for} to generate BCs tables for the user-selected
filters, and all values of {\tt ialf} in a single execution (i.e., the value
of {\tt ialf} that is read from selectbc.data is ignored at this stage).
Figure \ref{fig:figA2} shows how easy it is to generate the BC tables in the
case where {\tt selectbc.data} have the values given in the first seven lines of
Figure \ref{fig:figA1}. For the sake of this example, once
{\tt get4bctables.for} has been compiled, and the executable module has been
named {\tt get4bctables}, one simply has to enter {\tt get4bctables} (on a
Linux-like environment this holds true if the executable is placed on the
{\tt \$PATH}. Otherwise, all executables discussed in this Appendix can be
called from the directory they are located in by adding {\tt ./} to their
names e.g., {\tt ./get4bctables}).
This will generate a request for the $E(B-V)$ value of interest, to which we
have responded with a value of $0.035$ (from the permitted range $0.0 \le E(B-V)
\le 0.72$). The code then proceeds to retrieve from the library the data for
the selected filters (which are printed on the monitor for the convenience of
the user) and to generate \verb+bc_std.data+, \verb+bc_p04.data+,
\verb+bc_p00.data+, and \verb+bc_m04.data+ for the specified reddening.
These \verb+bc_***.data+ files can then be used e.g., to compute bolometric
corrections (and colour indices) of stars of known $\teff, \logg$ and $\feh$, at
the reddening entered above. This can be done using the program
{\tt bcstars.for} (which will call subroutines contained in {\tt bcutil.for}
for the appropriate value of {\tt ialf}).

The above procedure (editing {\tt selectbc.data}, executing
{\tt get4bctables}, and {\tt bcstars}) is easily repeated to obtain the BCs for
a different reddening and/or an alternative selection of filters and/or a
different list of input stellar parameters. Importantly, our programs
assume a certain directory structure to be in place. Hence, the libraries in
{\tt BCtables.tar.gz} and {\tt BCtables2.tar.gz} {\it must} be unpacked in the
same directory as the various FORTRAN codes and {\tt selectbc.data}.

As described above, our routines can be used to compute bolometric corrections
at given input of $\teff, \logg$ and $\feh$ for selected values of
[$\alpha$/Fe]. These inputs can be e.g., from a sample of stars; from a list
covering a range of $\teff$ for the sake of deriving synthetic colour-$\teff$
relations at various gravities and metallicities; or from theoretical
isochrones as discussed in more detail further below.

Since one of the main uses of a library of bolometric corrections is to
transform isochrones from the theoretical to various observational planes, we
have provided an example of a computer code that will automatically select, and
interpolate in, the appropriate {\verb+bc_***.data+} file for the value of
[$\alpha$/Fe] that was assumed in the computation of the stellar models.
This program, which has been given the name {\tt iso2cmd.for}, must be compiled
and linked with the interpolation subroutines in {\tt bcutil.for} in order to
produce an executable module.  Although it currently assumes that
the input isochrones are given in the format of the Victoria-Regina (hereafter
VR) models by \cite{vbf14}, those who wish to use the published grids
of other workers should find it easy to modify {\tt iso2cmd.for} to accommodate
the format in which they have been presented.

\begin{figure}
\begin{center}
\includegraphics[width=0.45\textwidth]{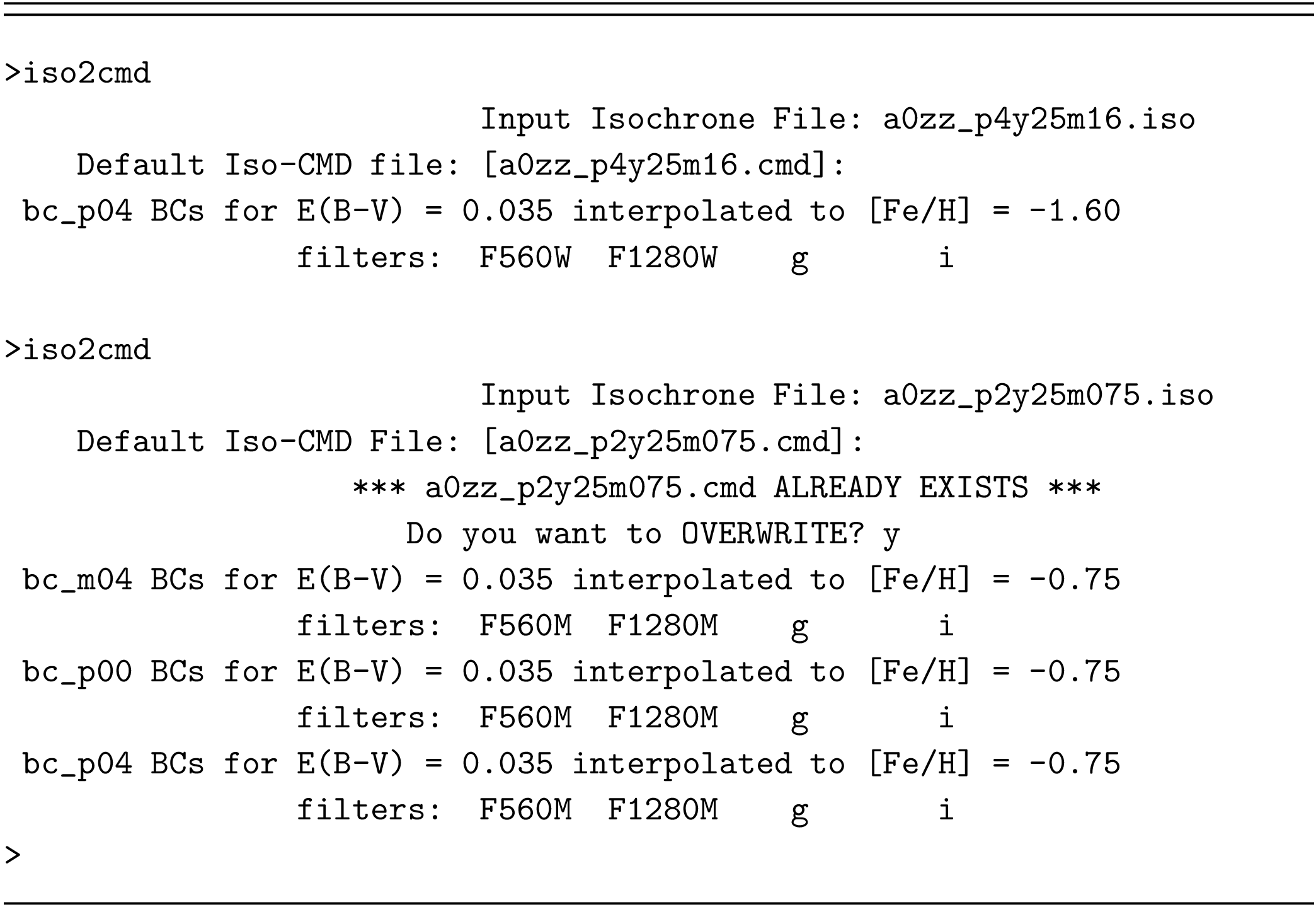}
\caption{Sample executions of {\tt iso2cmd} in which the BCs that were 
generated by {\tt get4bctables} (see the previous figure) have been applied
to two different isochrones.  These examples are explained in detail in the
text.}\label{fig:figA3}
\end{center}
\end{figure}

Suppose that the {\verb+bc_***.data+} files contain the BCs that were generated
for the case that is the subject of Figures \ref{fig:figA1} and
\ref{fig:figA2}.  The application of those
transformations to two different VR isochrones is illustrated in Figure
\ref{fig:figA3}.  If
the executable module has been given the name {\tt iso2cmd}, entering this name
at the prompt will generate a request for the name of the input isochrone.
In the first example, the name {\verb+a0zz_p4y25m16.iso+} has been entered
after the colon.  By convention \citep[see the Appendix in][]{vbf14},
{\tt a0} refers to the solar mix of heavy elements derived by \cite{ags09},
{\tt zz} specifies the entire group of $\alpha$-elements, and the rest of the
isochrone name indicates that [$\alpha$/Fe] $= +0.4$ ({\verb+_p4+}),
$Y = 0.25$ ({\tt Y25}), and [Fe/H] $= -1.6$ ({\tt m16}).  (Had the models been
computed for [$\alpha$/Fe] $ = -0.4$,
the first 7 characters of the isochrone name would have been {\verb+a0zz_m4+}.)
After the input isochrone has been identified, {\tt iso2cmd} suggests that the
output file be given the same name, but with the extension {\tt .cmd} (see the
third line of Figure \ref{fig:figA3}). If that name is satisfactory, the user
simply has to click on the ``enter'' key (without typing anything); otherwise,
a different name (e.g., {\tt testcmd.dat}) can be entered.  The last two lines
in the first example indicate that interpolations were carried out in
\verb+bc_p04.data+, which assume $E(B-V) = 0.035$, to obtain the
reddening-corrected BCs at [Fe/H] $= -1.60$ for the filters $F560W$, $F1280W$,
$g$, and $i$.

In the second example, an isochrone for [$\alpha$/Fe] $= 0.2$, $Y=0.25$, and
[Fe/H] $= -0.75$ (\verb+a0zz_p2y25m075.iso+) is similarly transformed to the
observed plane. If, as this in this case, a {\tt .cmd} file of the same name
already exists, the code issues a warning when the ``enter'' key is pressed to
accept the default name, and the user is given the option of overwriting that
file.  By entering {\tt y} (or {\tt yes}), that option is accepted.  (Had
{\tt n} or {\tt no} been entered, the user would have been asked to provide an
alternative name for the output file.)  Because the value of [$\alpha$/Fe] does
not lie on the ``standard'' relation between [$\alpha$/Fe] and [Fe/H], it is
necessary to perform interpolations in the \verb+bc_m04.data+,
\verb+bc_p00.data+, \verb+bc_p04.data+ tables. Doing so yields the BCs for
[$\alpha$/Fe] $ = -0.4$, 0.0, and $+0.4$, at [Fe/H] $= -0.75$;  consequently,
a further interpolation is carried out in these results to obtain the
transformations at [$\alpha$/Fe] $= +0.2$ at the same metallicity.

\begin{figure}
\begin{center}
\includegraphics[width=0.45\textwidth]{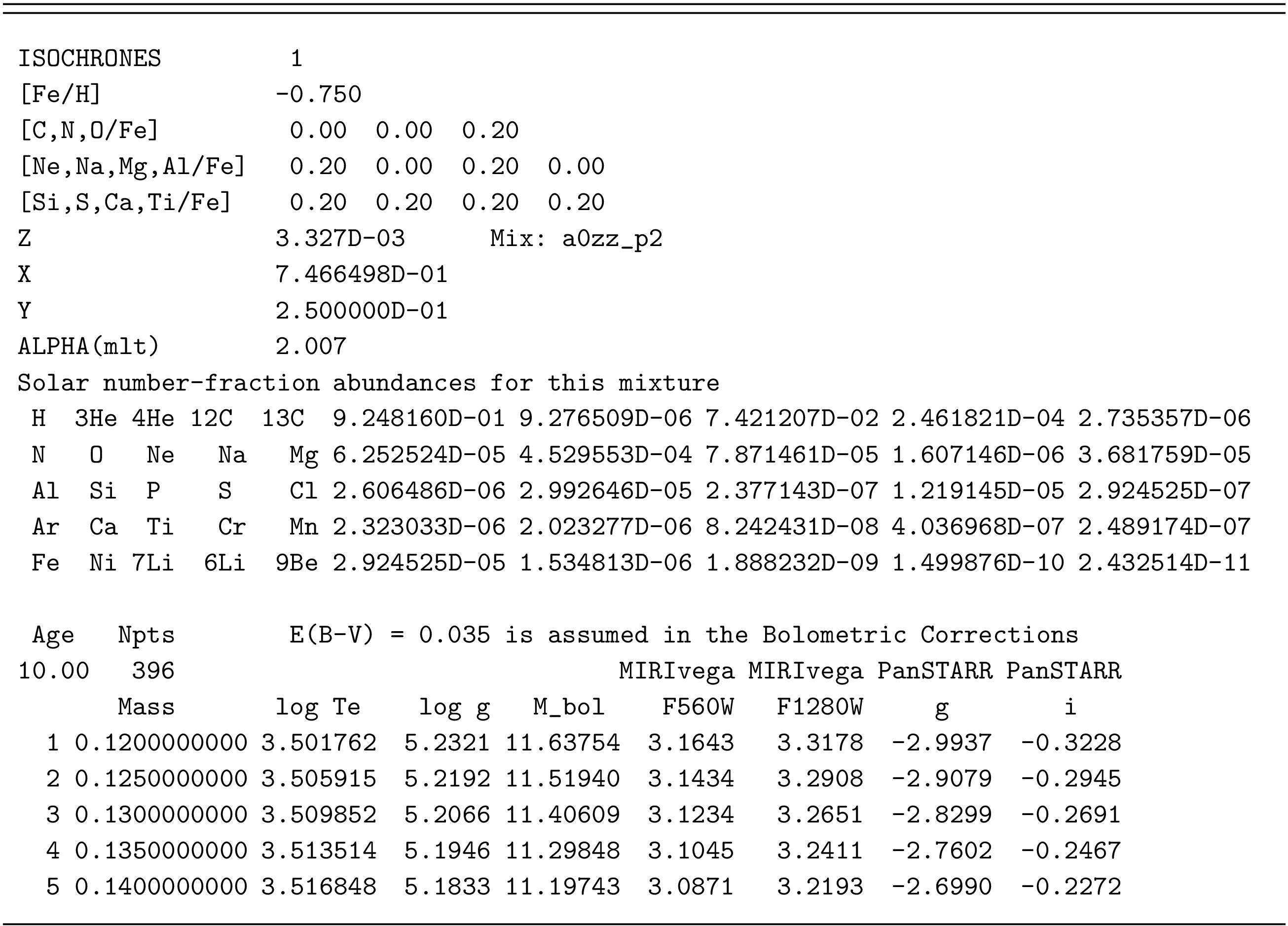}
\caption{The first 24 lines in the output {\tt .cmd} file of the [Fe/H]
$=-0.75$ isochrone that was considered in the previous figure. The text
contains a brief explanation of the information that is
listed.}\label{fig:figA4}
\end{center}
\end{figure}

Figure \ref{fig:figA4} reproduces the first two dozen lines in the
\verb+a0zz_p2y25m075.cmd+ file that would have been generated in the second of
the two test cases just described.  The first 9 of the header lines list the
number of isochrones, the assumed [Fe/H] value and the initial [$m$/Fe] ratios
of the most important light elements, the initial mass-fraction abundances of
the metals ($Z$), hydrogen ($X$), and helium ($Y$), the name of the metals
mixture (\verb+a0zz_p2+), and the solar-calibrated value of the mixing-length
parameter.  The next 6 lines give the solar number-fraction abundances that
are used by the Victoria stellar structure code to evaluate the surface
[$m$/H] and [$m$/Fe] of the most abundant metals along evolutionary tracks.
They are followed by the listing of the properties of individual models along
a $10.0$ Gyr isochrone (specifically, the mass in solar units, $\log\,\teff$,
$\log\,g$, $M_{\rm bol}$), together with the reddening-corrected BCs for the
indicated value of $E(B-V)$ and the 4 filters that are identified.

The absolute magnitude in any filter $\zeta$ can be determined from $M_\zeta =
M_{\rm bol} - BC_\zeta = -2.5 \log\frac{L}{L_\odot} + M_{\rm{Bol},\odot} - BC_\zeta$,
where $L$ is the luminosity at a given $\teff, \logg$ and $\feh$, and we refer
to Section \ref{sec:phy} for a discussion on the adoption of solar bolometric
magnitude and luminosity. Differences in absolute magnitudes, $M_\zeta -M_\eta$,
or equivalently, differences in the relevant bolometric corrections,
$BC_\eta -BC_\zeta$
(note the reversal in the subscripts), then correspond to the predicted colour
index $\zeta - \eta$. Although the header lines are specific to the VR
computations, users should find it easy to produce equivalent listings (lines
17--) of  isochrones produced by other workers with no more than a small
amount of editing of {\tt iso2cmd.for}.

\section[]{Extinction coefficients}\label{appB}

As discussed in length in Paper I, in presence of a given extinction the
colour excess varies with stellar spectral type. This effect can be quite
dramatic,
and our interpolation routines allow for its exact treatment. However, there
can be instances where for practical purposes the assumption of an
average extinction coefficients is sufficient, or where the variation of
extinction coefficients can be described with sufficient accuracy as function
of $\teff$ and $\feh$. Table \ref{tab:tabler} provides extinction coefficients
for the photometric systems studied in this paper, determined in the same
manner as those reported in the analogous table included in Paper I.

\begin{table*}
\centering
\caption{Extinction coefficients relevant to turnoff stars with
  $\teff \lesssim 7000\,{\rm{K}}^{a}$. Both mean extinction coefficients
  $\langle R_\zeta\rangle$, along with a linear fit of $R_\zeta$ as function of 
  effective temperature and metallicity are provided.}\label{tab:tabler}
\smallskip
%\begin{scriptsize}
\begin{tabular}{lcccccclccccc}
\hline
\hline 
\noalign{\smallskip}
  & & \multicolumn{4}{c}{$R_\zeta = a_0 + T_4\,(a_1 + a_2\,T_4) + a_3\,$[Fe/H]} & & & & \multicolumn{4}{c}{$R_\zeta = a_0 + T_4\,(a_1 + a_2\,T_4) + a_3\,$[Fe/H]}\\
Filter & $\langle R_\zeta\rangle$ & \multispan4\hrulefill &  \phantom{~~~~~~~~~~} & Filter & $\langle R_\zeta\rangle$ & \multispan4\hrulefill \\
 & & $a_0$ & $a_1$ & $a_2$ & $a_3$ & &  & & $a_0$ & $a_1$ & $a_2$ & $a_3$ \\
\noalign{\smallskip}
\hline
\noalign{\smallskip}
\noalign{\centerline{\tt \emph{JWST} - MIRI}} & & & \\
%\noalign{\vskip -0.23in} \\
 $F560W$ & 0.102 & 0.1061 & $-0.0132$    & \phantom{+}$0.0102$ & --- & &
 $F770W$ & 0.102 & 0.0986 & \phantom{+}$0.0106$  & $-0.0088$   & --- \\
 $F1000W$ & 0.102 & 0.0964 & \phantom{+}$0.0170$ & $-0.0135$   & --- & &
 $F1130W$ & 0.102 & 0.0975 & \phantom{+}$0.0140$ & $-0.0115$   & --- \\
 $F1280W$ & 0.102 & 0.1081 & $-0.0204$   & \phantom{+}$0.0166$ & --- & &
 $F1500W$ & 0.102 & 0.1073 & $-0.0171$   & \phantom{+}$0.0134$ & --- \\
\noalign{\vskip -0.1in} \\
\noalign{\centerline{\tt \emph{JWST} - NIRCam}} \\
%\noalign{\vskip -0.23in} \\
 $F070W$ & 2.314 & 2.2385 & \phantom{+}$0.1738$ & $-0.0803$ & \phantom{+}$0.0010$  & &
 $F090W$ & 1.514 & 1.4447 & \phantom{+}$0.1833$ & $-0.1125$ & ---  \\
 $F115W$ & 1.011 & 0.9910 & \phantom{+}$0.0313$ & \phantom{+}$0.0018$ & ---  & &
 $F140M$ & 0.727 & 0.7233 & \phantom{+}$0.0051$ & \phantom{+}$0.0011$ & ---  \\
 $F150W2$ & 0.700 & 0.4673 & \phantom{+}$0.5534$ & $-0.2778$ & \phantom{+}$0.0023$ & &
 $F150W$  & 0.663 & 0.6425 & \phantom{+}$0.0454$ & $-0.0189$ & \phantom{+}$0.0006$ \\
 $F162M$  & 0.574 & 0.5719 & \phantom{+}$0.0032$ & --- & --- & &
 $F182M$  & 0.471 & 0.4709 & \phantom{+}$0.0009$ & $-0.0007$ & --- \\
 $F200W$  & 0.425 & 0.4159 & \phantom{+}$0.0261$ & $-0.0195$ & --- & &
 $F210M$  & 0.382 & 0.3747 & \phantom{+}$0.0260$ & $-0.0218$ & --- \\
 $F250M$  & 0.287 & 0.2828 & \phantom{+}$0.0131$ & $-0.0109$ & --- & &
 $F277W$  & 0.253 & 0.2554 & $-0.0086$ & \phantom{+}$0.0085$ & --- \\
 $F300M$  & 0.216 & 0.2222 & $-0.0175$ & \phantom{+}$0.0131$ & --- & &
 $F322W2$ & 0.214 & 0.2059 & \phantom{+}$0.0254$ & $-0.0189$ & --- \\
 $F335M$  & 0.179 & 0.1774 & \phantom{+}$0.0056$ & $-0.0045$ & --- & &
 $F356W$  & 0.166 & 0.1699 & $-0.0102$ & \phantom{+}$0.0075$ & --- \\
 $F360M$  & 0.158 & 0.1597 & $-0.0047$ & \phantom{+}$0.0036$ & --- & &
 $F410M$  & 0.132 & 0.1349 & $-0.0108$ & \phantom{+}$0.0092$ & --- \\
 $F430M$  & 0.121 & 0.1313 & $-0.0335$ & \phantom{+}$0.0267$ & --- & &
 $F444W$  & 0.119 & 0.1270 & $-0.0246$ & \phantom{+}$0.0200$ & --- \\
 $F460M$  & 0.106 & 0.1103 & $-0.0129$ & \phantom{+}$0.0106$ & --- & & 
 $F480M$  & 0.101 & 0.0987 & \phantom{+}$0.0083$ & $-0.0075$ & --- \\
\noalign{\vskip -0.1in} \\
\noalign{\centerline{\tt SkyMapper}} \\
%\noalign{\vskip -0.23in} \\
 \phantom{F4}$u$ & 4.900 & 3.3743 & \phantom{+}$4.5098$ & $-3.2967$ & $-0.0193$ & &
 \phantom{F4}$v$ & 4.550 & 4.3395 & \phantom{+}$0.7243$ & $-0.6196$ & $-0.0028$ \\
 \phantom{F4}$g$ & 3.446 & 2.9349 & \phantom{+}$1.2782$ & $-0.7275$ & $-0.0054$ & &
 \phantom{F4}$r$ & 2.734 & 2.6011 & \phantom{+}$0.2952$ & $-0.1284$ & --- \\
 \phantom{F4}$i$ & 1.995 & 1.9686 & \phantom{+}$0.0394$ & \phantom{+}$0.0069$ & --- & &
 \phantom{F4}$z$ & 1.468 & 1.3831 & \phantom{+}$0.2551$ & $-0.1886$ & --- \\
\noalign{\vskip -0.1in} \\
\noalign{\centerline{\tt Tycho}} \\
%\noalign{\vskip -0.23in} \\
 \phantom{F4}$H_P$     & 3.239 & 2.0611 & \phantom{+}$2.9605$ & $-1.6990$ & $-0.0133$ & & 
 \phantom{F4}$B_T$ & 4.222 & 3.6609 & \phantom{+}$1.6185$ & $-1.1570$ & $-0.0126$ \\
 \phantom{F4}$V_T$ & 3.272 & 3.0417 & \phantom{+}$0.5745$ & $-0.3231$ & $-0.0015$ & &
               &       &        &              &           &           \\
\noalign{\vskip -0.1in} \\
\noalign{\centerline{\tt Pan-STARRS1}} \\
%\noalign{\vskip -0.23in} \\
 \phantom{F4}$g$  & 3.666 & 3.1864 & \phantom{+}$1.2310$ & $-0.7350$ & $-0.0060$ & &
 \phantom{F4}$r$  & 2.709 & 2.6161 & \phantom{+}$0.2086$ & $-0.0928$ & --- \\
 \phantom{F4}$i$  & 2.108 & 2.0702 & \phantom{+}$0.0806$ & $-0.0291$ & --- & &
 \phantom{F4}$z$  & 1.609 & 1.5920 & \phantom{+}$0.0421$ & $-0.0236$ & --- \\
 \phantom{F4}$y$  & 1.338 & 1.3078 & \phantom{+}$0.0959$ & $-0.0759$ & --- & &
 \phantom{F4}$w$  & 2.824 & 1.9382 & \phantom{+}$1.9907$ & $-0.8899$ & $-0.0053$ \\
 \phantom{F4}$o$  & 2.594 & 1.3963 & \phantom{+}$2.6716$ & $-1.1671$ & $-0.0049$ & &
                  &       &        &                     &           &  \\
\noalign{\smallskip}
\hline
\noalign{\smallskip}
\end{tabular}
%\end{scriptsize}
\begin{minipage}{1\textwidth}
$^{a}$~Based on the differences in the bolometric corrections for 
  $E(B-V) = 0.0$ and $0.10$, assuming $\logg = 4.1$,
  $5250 \le \teff \le 7000$~K, and $-2.0 \le$ [Fe/H] $\le +0.25$, with 
  [$\alpha$/Fe] $=-0.4, 0.0$ and $0.4$ at each [Fe/H] value. Note that (i)
  $T_4 = 10^{-4}\,T_{\rm eff}$ in the fitting equation for $R_\zeta$, and (ii)
  for a nominal $E(B-V)$, the excess in any given $\zeta-\eta$ colour is
  $E(\zeta-\eta)=(R_\zeta-R_\eta)E(B-V)$, while the attenuation for a magnitude
  $m_\zeta$ is $R_\zeta E(B-V)$. Also see the discussion in Paper I. \\
\phantom{~~~~~~~~~~~~~~~~~~~~~~~}\\
\phantom{~~~~~~~~~~~~~~~~~~~~~~~}
\end{minipage}

\end{table*}

\end{document}